\documentclass[]{aa}
\usepackage{graphicx,amsmath}
\usepackage{txfonts}

\newcommand{\abin}{a_{\rm bin}}
\newcommand{\ebin}{e_{\rm bin}}
\newcommand{\Tbin}{T_{\rm bin}}

\newcommand{\cs}{c_{\rm s}}

\newcommand{\tirr}{T_{\rm irr}}
\newcommand{\tcool}{t_{\rm cool}}
\newcommand{\taueff}{\tau_{\rm eff}}
\newcommand{\tauu}{\tau_{\rm upper}}
\newcommand{\taul}{\tau_{\rm lower}}
\newcommand{\qcool}{{\cal Q}_{\rm cool}}
\newcommand{\rin}{R_{\rm in}}
\newcommand{\rout}{R_{\rm out}}
\newcommand{\agap}{a_{\rm gap}}
\newcommand{\egap}{e_{\rm gap}}

\begin{document}
\title{Three-dimensional evolution of radiative circumbinary discs: the size and shape of the inner cavity }
\titlerunning{3D evolution of radiative circumbinary discs}
\author{Arnaud Pierens 
 \inst{1}
 \and
  Richard P. Nelson\inst{2}}
\institute{ Laboratoire d'astrophysique de Bordeaux, Univ. Bordeaux, CNRS, B18N, all\'ee Geoffroy Saint-Hilaire, 33615 Pessac, France\\
\email{arnaud.pierens@u-bordeaux.fr}
\and
 Astronomy Unit, Queen Mary University of London, Mile End Road, London, E1 4NS, UK}

\abstract{
The  evolution of circumbinary discs and planets is often studied using two-dimensional (2D) numerical simulations, although recent work suggests that 3D effects may significantly alter the structure of the inner cavity created by the binary. \\
In this study, we present the results of 3D hydrodynamical simulations of  circumbinary discs that orbit around analogues of the Kepler-16 and Kepler-34 systems, including the effect of stellar heating  and radiative cooling on the thermal disc structure. We find that compared to their 2D counterparts, the structures of the cavities in 3D circumbinary disc models appear to reach a quasi-stationary state more rapidly, and in a subset of our runs the evidence for this is unambiguous. Furthermore, the sizes and eccentricities of the inner cavity are smaller in 3D compared to 2D. We attribute this difference to enhanced spiral wave dissipation in disc regions above the midplane, where the cooling time is of the order of the dynamical timescale, resulting in smaller inner cavity sizes in 3D disc models.\\
 Our results suggest that migrating planets should park closer to the central binary in 3D models of circumbinary discs, and point to the importance of including the 3D structure when simulating circumbinary discs and planets.
}

\maketitle

\section{Introduction}
To date, 15 circumbinary planets have been discovered around eclipsing binary systems consisting of main sequence stars. Most of these, such as Kepler-16b and Kepler-34b, were detected by the Kepler spacecraft (Doyle et al. 2011; Welsh et al. 2012), and recently TESS has also detected two circumbinary planets TOI-1338b (Kostov et al. 2020) and TIC-172900988 (Kostov et al. 2021). The first detection of a circumbinary planet by the radial velocity method recently confirmed that Kepler-16b has a mass similar to that of Saturn (Triaud et al. 2021), in agreement with the mass estimate obtained by fitting a photo-dynamical model to the transit data (Doyle et al. 2011).

The majority (10 out of 15) of these circumbinary planets orbit just outside of the dynamical instability zone (Holman \& Wiegert 1999), where perturbations due to the binary would cause planets to be ejected. Simulations show that the structure of the circumbinary disc would be highly perturbed in this region during the planet formation epoch, and numerous studies have shown that the formation of a significantly eccentric precessing cavity makes it highly unlikely that in situ planet formation via planetesimal accretion can occur there (Marzari et al. 2008; Lines et al. 2016). Futhermore, recent work has shown that the inner eccentric regions of circumbinary discs are prone to a parametric instability that generates hydrodynamical turbulence  (Papaloizou 2005; Barker \& Ogilvie 2014), whose impact on planet formation is twofold (Pierens et al. 2020, 2021): i) It makes the collision velocities between small particles too high to allow grain growth, such that forming a massive planetesimal seed for pebble accretion becomes difficult;  ii) It stirs up pebbles and renders pebble accretion very inefficient. Hence, forming circumbinary planets in situ near the dynamical instability zone via pebble accretion also appears to be very challenging.  Instead, circumbinary planets could form at large distances from the binary,  where dust growth and pebble accretion are more efficient, and then migrate in before stalling at their currently observed locations. 

There is a substantial body of work that has examined the migration of planets in circumbinary discs, and a general picture that has emerged is that a planet embedded in a circumbinary disc migrates to the inner edge of the disc cavity. For moderate binary eccentricities, migration stalls at this location due to a positive corotation torque being exerted on the planet that counteracts the Lindblad torque  (Masset et al. 2006; Pierens \& Nelson 2007; Kley \& Haghighipour 2014; Penzlin et al. 2021). For binaries with higher eccentricities, the planet can acquire significant eccentricity due to its interaction with the highly eccentric circumbinary disc, which makes migration stall further away from the central binary, at distances generally beyond the observed locations (Pierens \& Nelson 2013; Kley \& Haghighipour 2015). This indicates that models in  which the disc eccentricity remains small are generally better for providing a good fit of the observed orbital parameters. For example, Mutter et al. (2017) and Penzlin et al. (2021) recently found that a  migrating planet with mass sufficient to open a partial gap tends to circularize  the inner cavity and subsequently  migrates closer to the central binary. In a recent paper, Coleman et al. (2022)  arrived to the same conclusion in a scenario where the disc is circularised as a result of the effect of the dust feedback onto the gas. 

In this paper we examine whether or not 3D rather 2D hydrodynamical simulations may also lead to better agreement between simulated circumbinary systems and observations. Our motivation stems from the fact that in our previous simulations of Kepler-413 analogues (Pierens \& Nelson 2018), we found that the density peak associated with the outer edge of the inner cavity was located at a similar location to the circumbinary planet Kepler-413b when simulated in 3-dimensions, whereas the peak was located further out in 2D simulations.  In comparison with our previous works,  in this paper we have significantly improved the modelling of the disc thermodynamics  by incorporating irradiation heating from the central binary. This is important since it has been shown that accurate treatment of the thermodynamics inside the cavity has a significant impact on modelling circumbinary discs (Sudarshan et al. 2022).  Although it remains uncertain to what degree some of our simulations have converged to a steady state, because of the computational expense of running 3D simulations, our results suggest that 3D models of circumbinary discs which include irradiation and cooling reach a quasi-steady state much earlier than 2D models, and that both the size and eccentricity of the inner cavity are smaller in 3D compared to 2D.

This paper is organised as follows. In Sect.~2 and 3, we describe the physical and numerical models respectively In Sect.4, we present the results of our 3D and 2D simulations of circumbinary discs around binaries with parameters equivalent to Kepler-16(AB) and Kepler-34(AB). We discuss our results in Sect.~5 and summarize our main findings in Sect.\-6.

\begin{figure}
\centering
\includegraphics[width=\columnwidth]{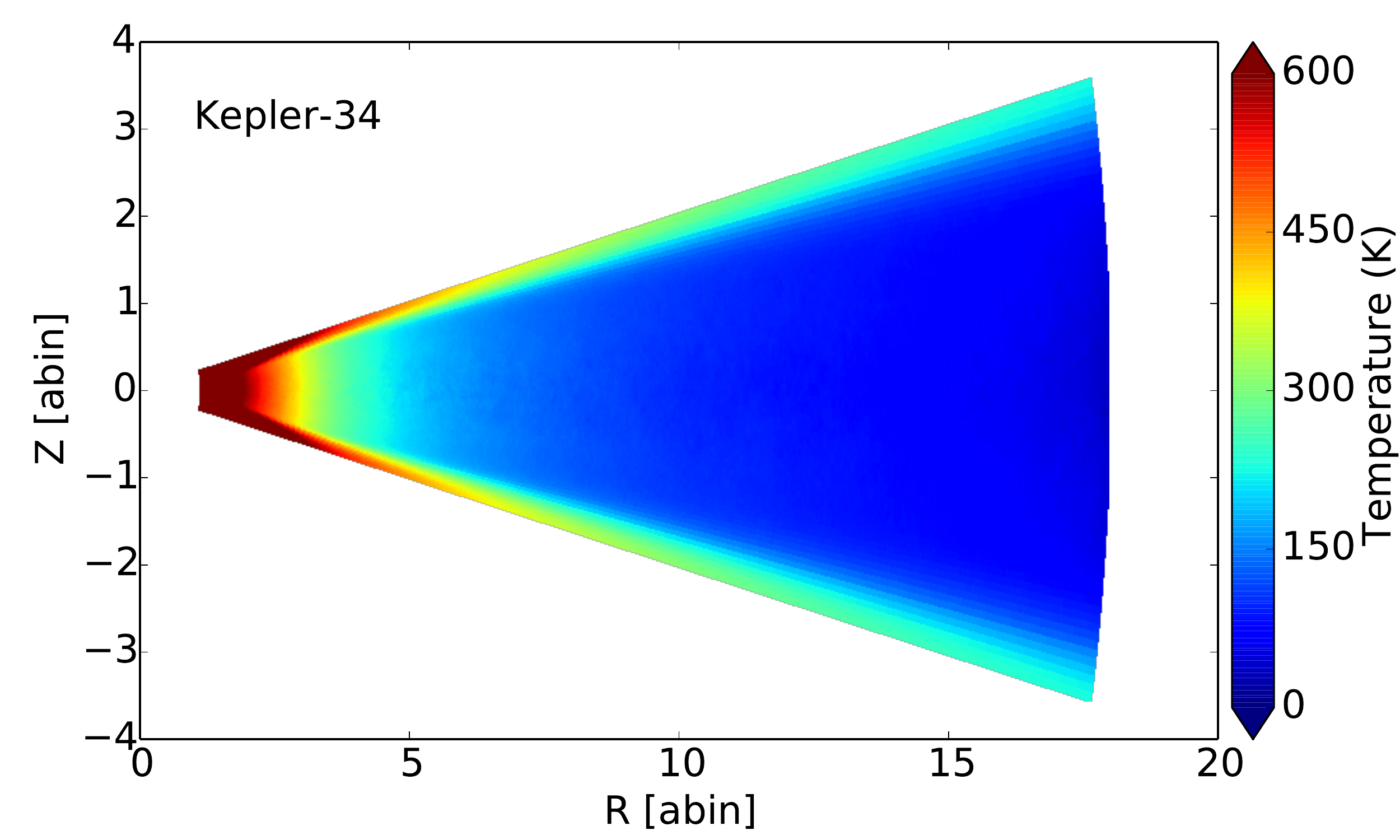}
\includegraphics[width=\columnwidth]{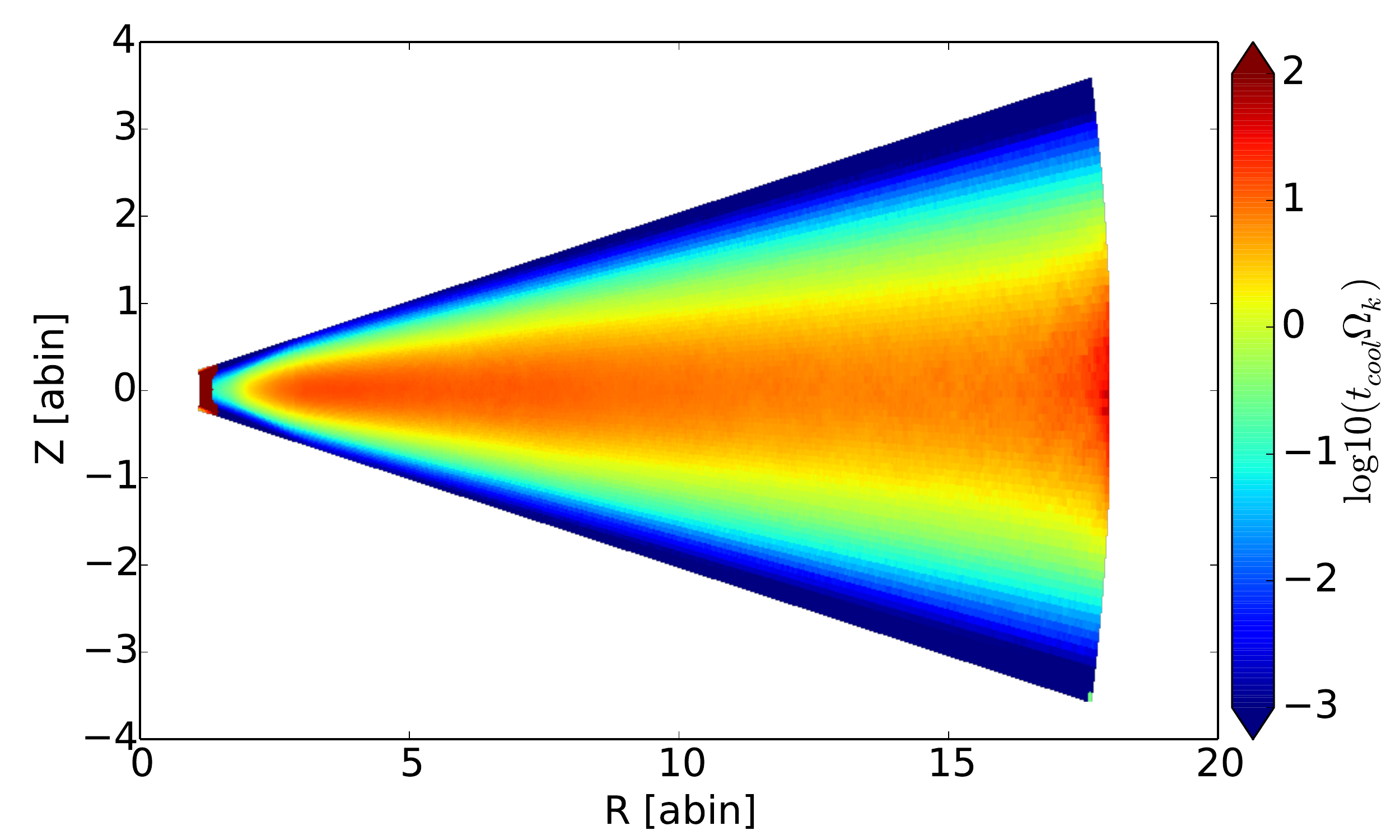}
\caption{Top: for Kepler-34, two-dimensional R-Z temperature distribution  for the initial disc model. Bottom: two-dimensional R-Z distribution of the dimensionless cooling time $\beta=\tcool \Omega_k$}
\label{fig:temp0_k34}
\end{figure}

\begin{figure}
\centering
\includegraphics[width=\columnwidth]{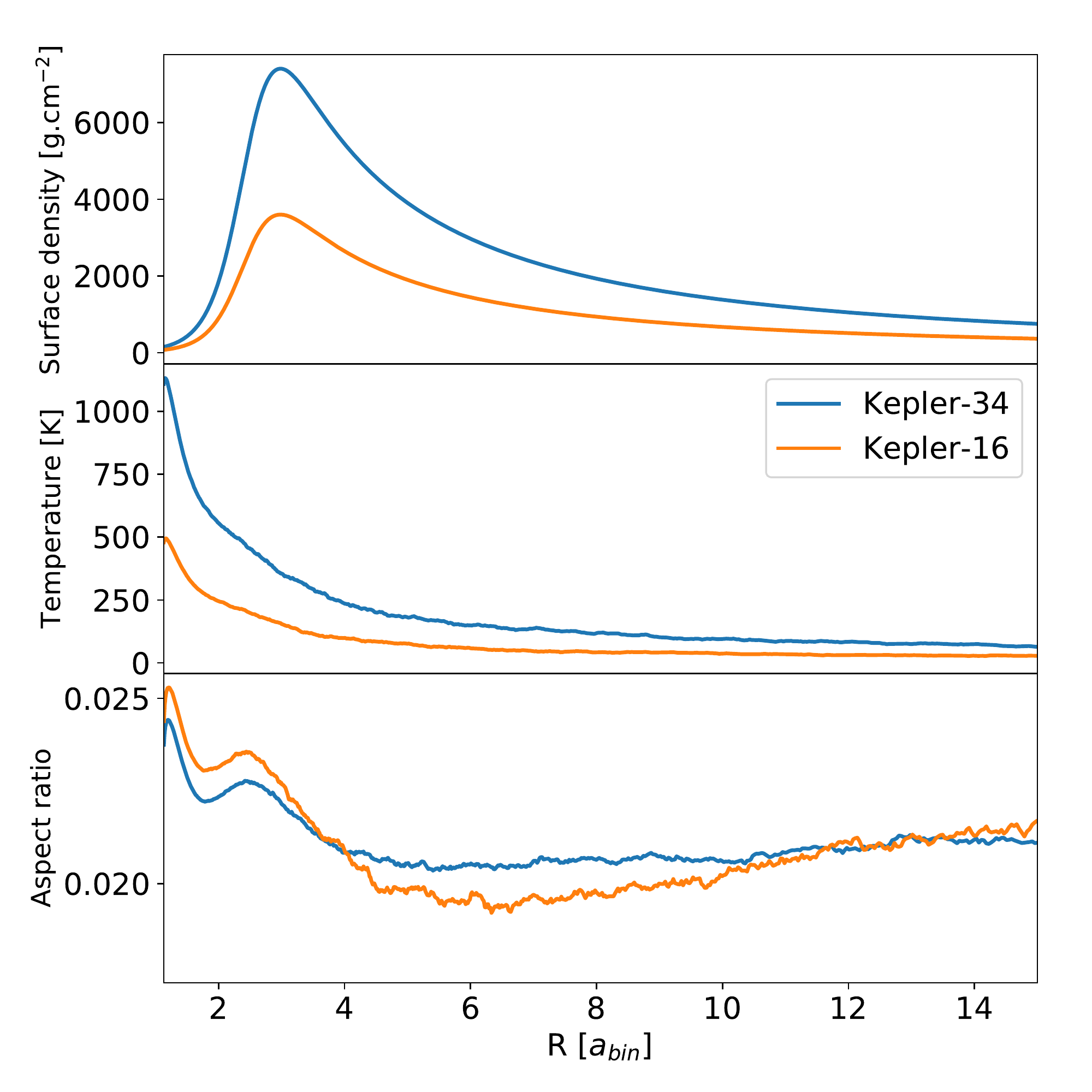}
\caption{Initial radial profiles of the surface density (top panel), temperature (middle panel), and aspect ratio (bottom panel).}
\label{fig:disc0}
\end{figure}

\begin{figure*}
\centering
\includegraphics[width=0.8\textwidth]{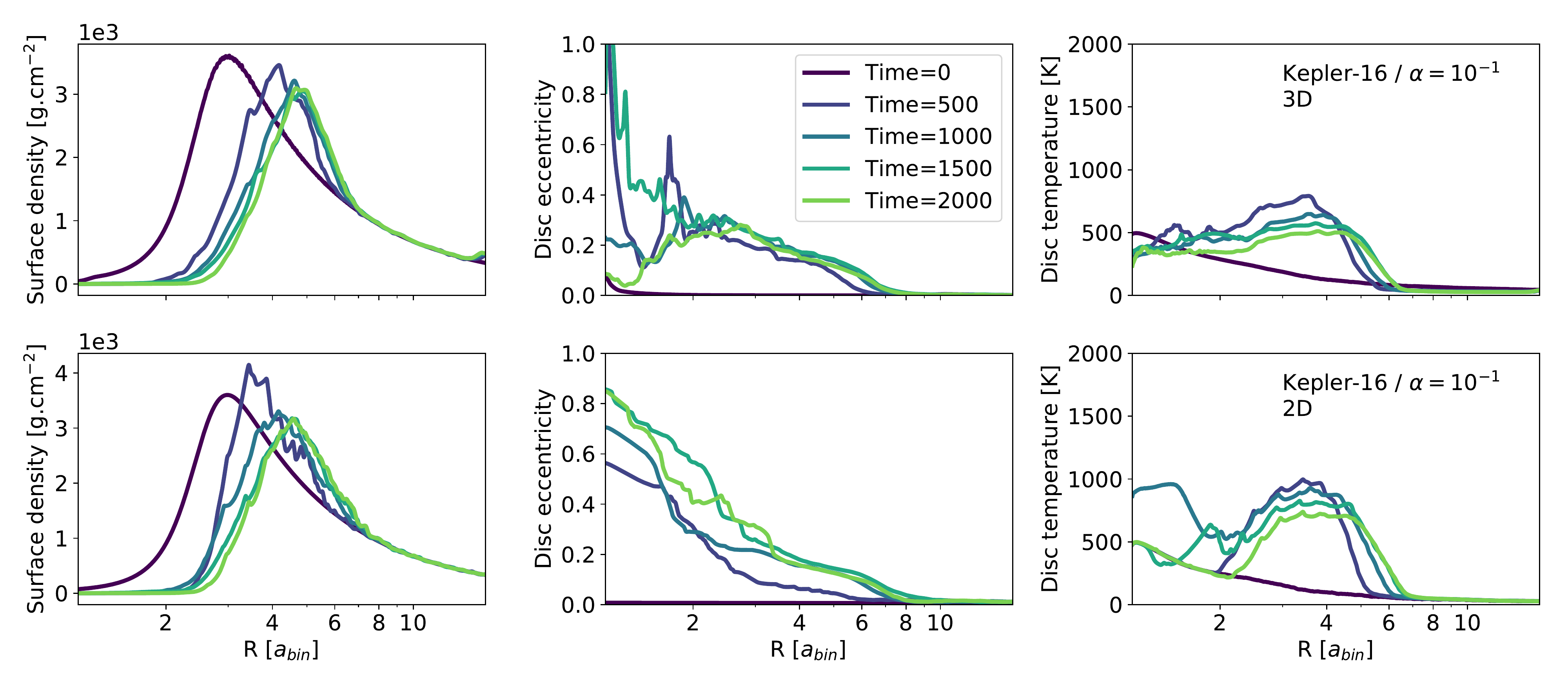}
\includegraphics[width=0.8\textwidth]{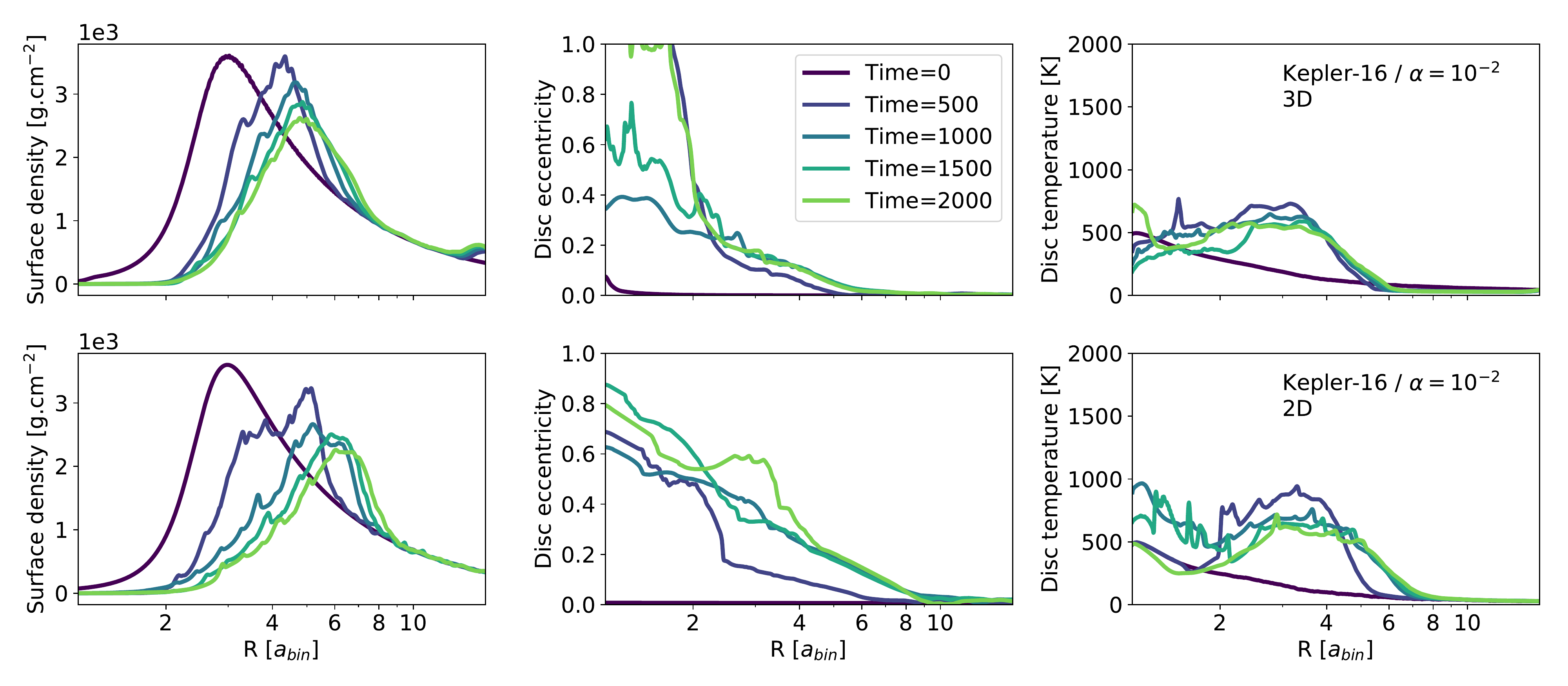}
\includegraphics[width=0.8\textwidth]{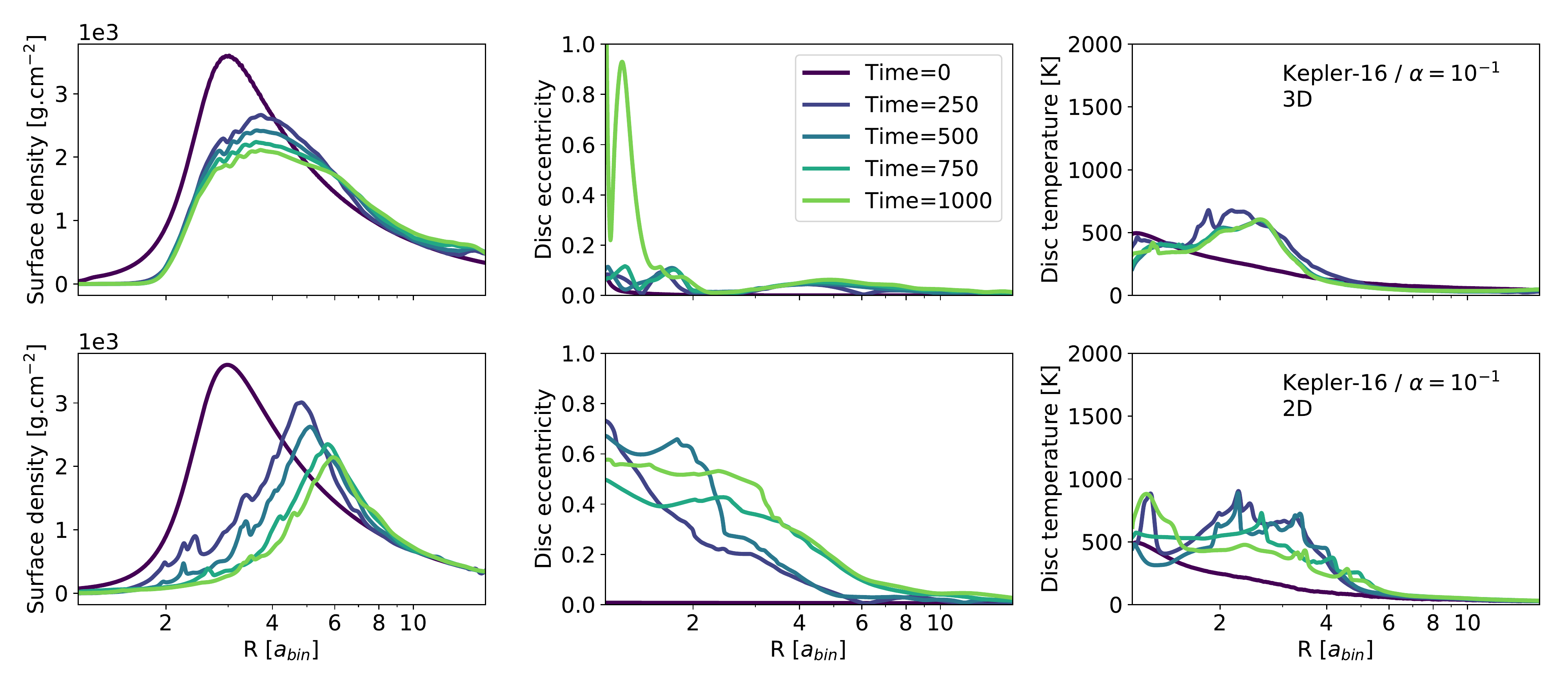}
\caption{Azimuthally-averaged surface density (left panels), disc eccentricity (middle panels), and temperature (right panels) for the 2D and 3D Kepler-16 analogues.}
\label{fig:kep16_1}
\end{figure*}

\begin{figure}
\centering
\includegraphics[width=\columnwidth]{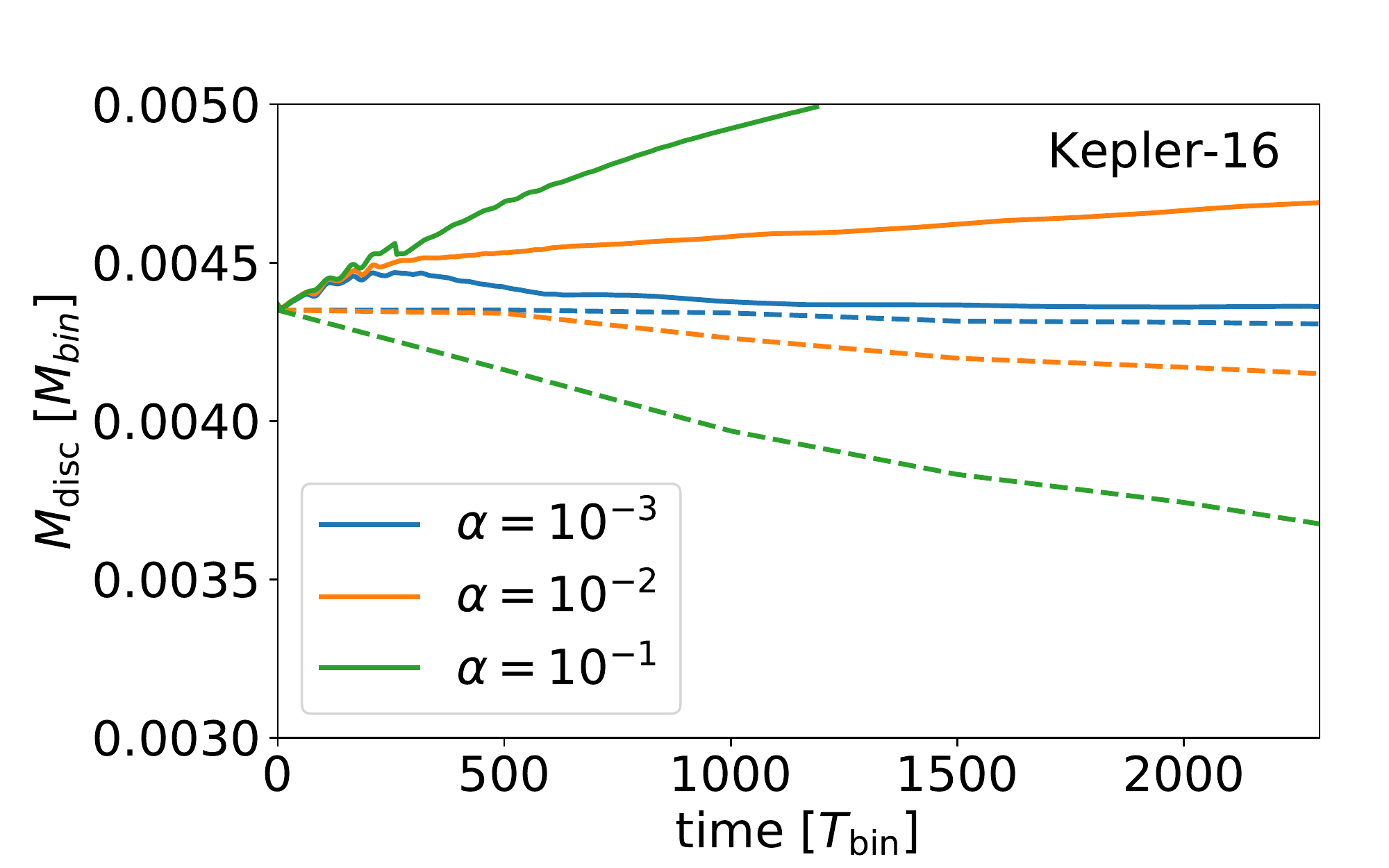}
\caption{Temporal evolution of the disc mass  for the Kepler-16 simulations. Solid  (resp. dashed)  lines correspond to 3D (resp. 2D) models.}
\label{fig:discmass_k16}
\end{figure}

\begin{figure*}
\centering
\includegraphics[width=\textwidth]{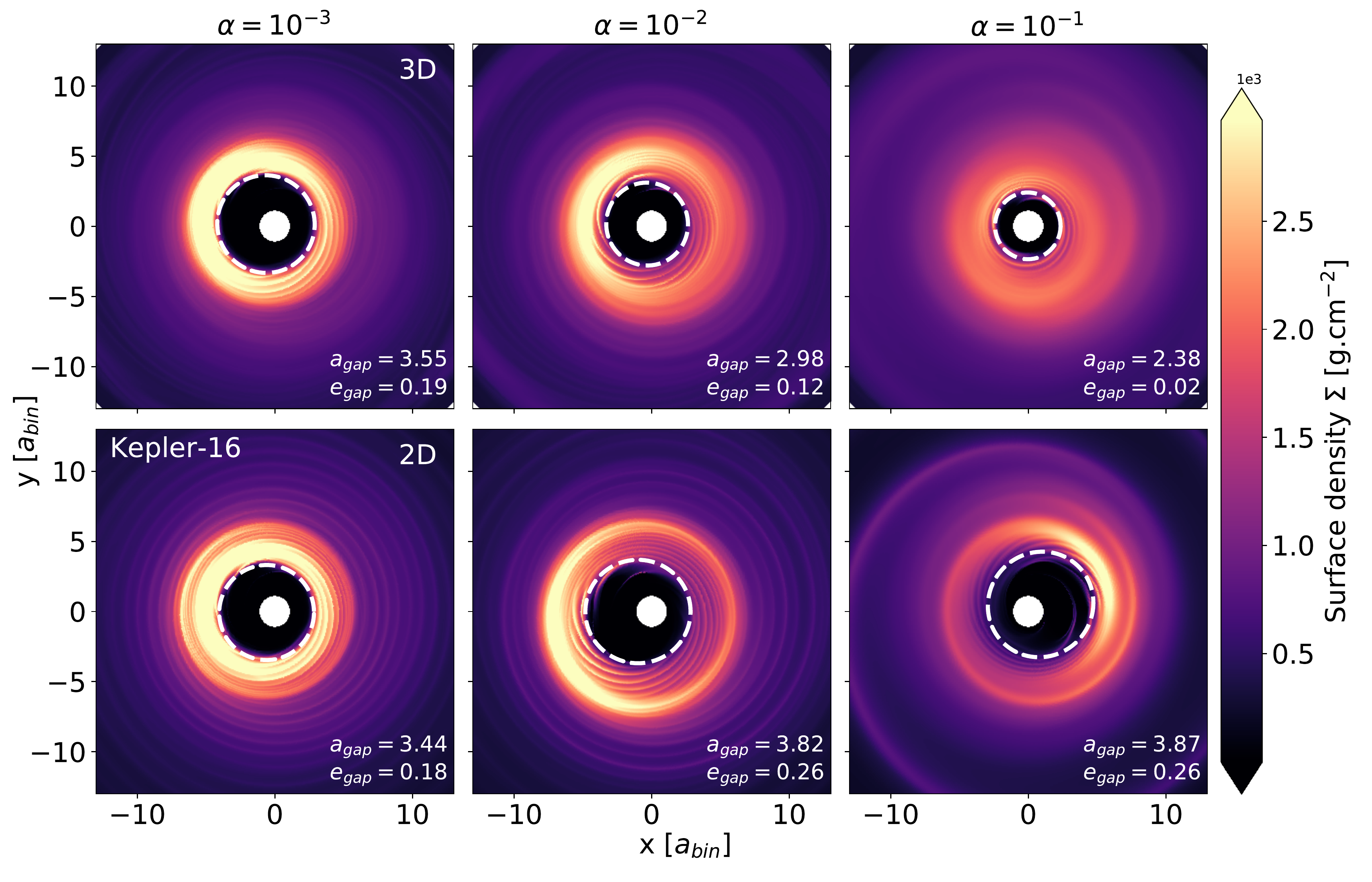}
\caption{Surface density distributions for the 2D and 3D Kepler-16 analogues for different values of the $\alpha$ parameter. The snapshots correspond to simulation runs times of  $2000$ $\Tbin$ when $\alpha=10^{-3}$ and $10^{-2}$, and $1500\Tbin$ when $\alpha=0.1$.}
\label{fig:2d_kepler16}
\end{figure*}

\begin{figure}
\centering
\includegraphics[width=\columnwidth]{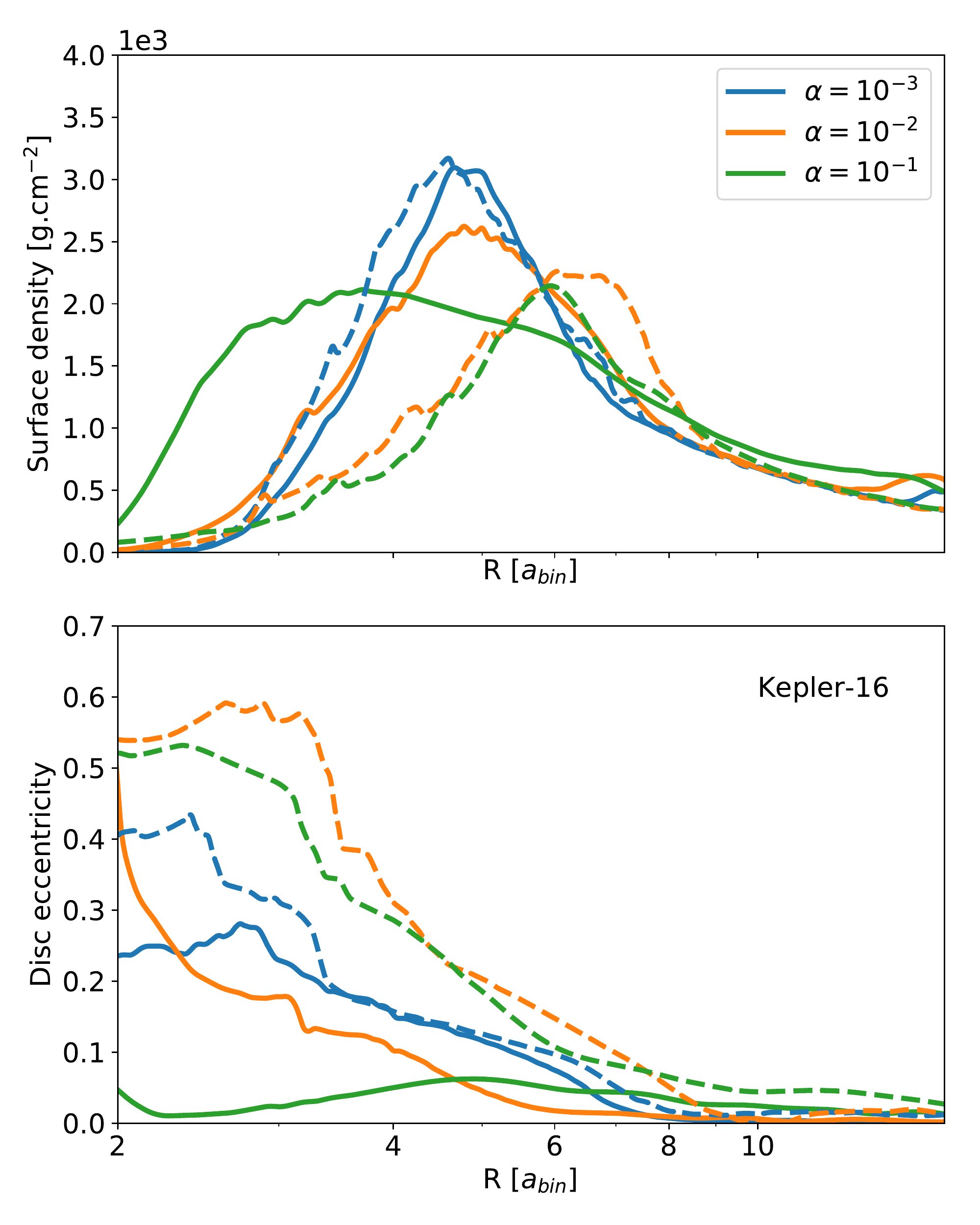}
\caption{For Kepler-16 and for each value of $\alpha$ we considered, comparison of the surface density (top panel) and eccentricity (bottom panel) profiles  obtained in the 3D (solid lines) and 2D (dashed lines) models.}
\label{fig:profiles_k16_t1000}
\end{figure}

\begin{figure*}
\centering
\includegraphics[width=\textwidth]{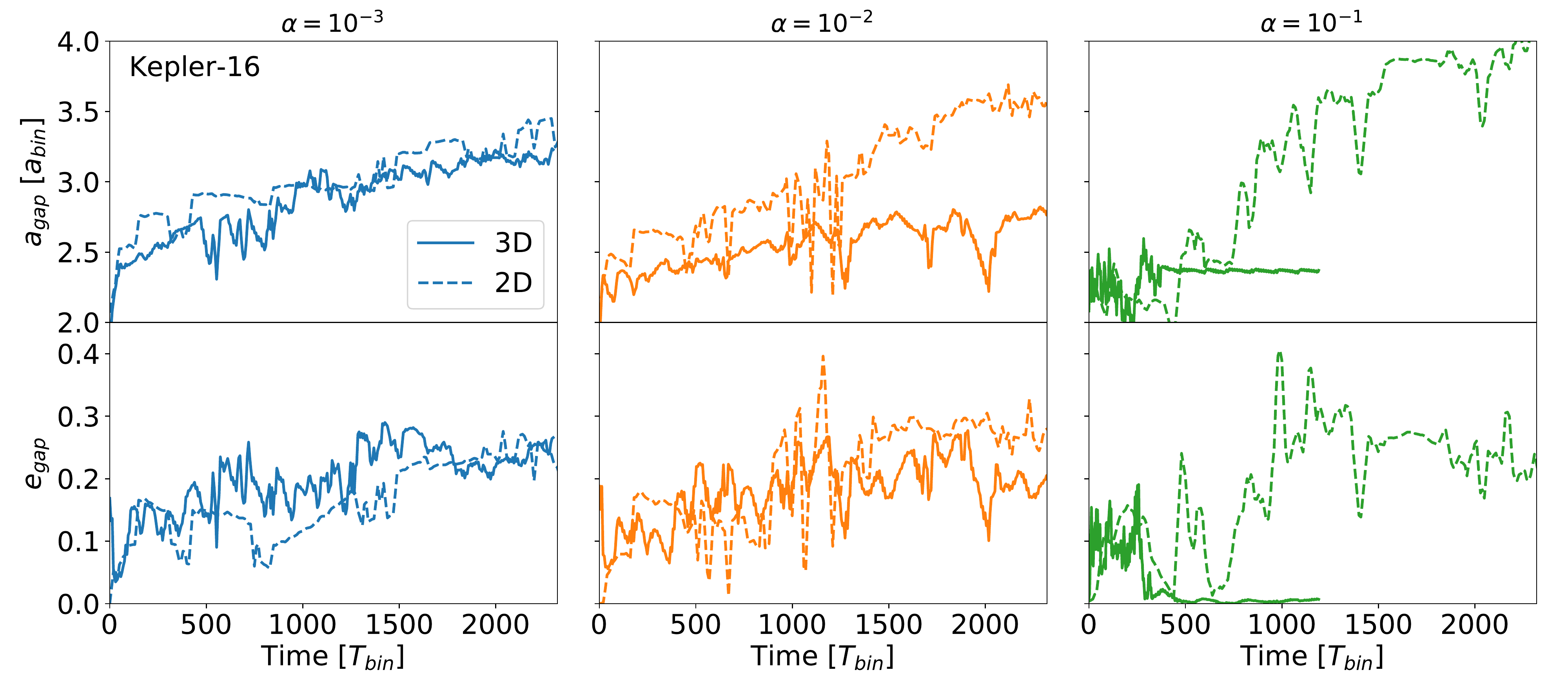}
\caption{Time evolution of the inner cavity semi-major axis $\agap$ and eccentricity $\egap$ for each Kepler-16 analogue simulation.}
\label{fig:aegap_k16}
\end{figure*}

\begin{figure}
\centering
\includegraphics[width=\columnwidth]{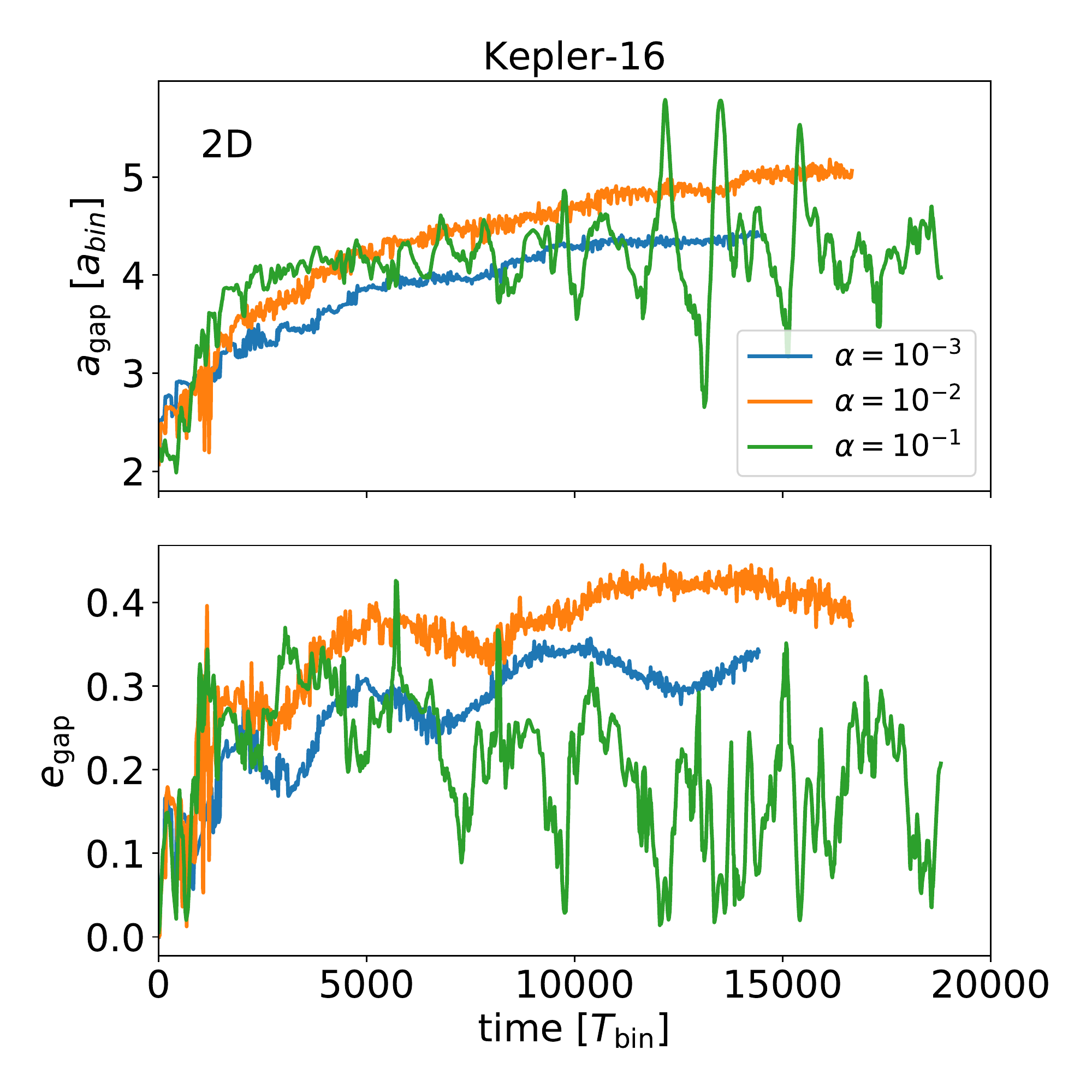}
\caption{Time evolution, over longer run times of $2\times 10^4$ binary orbits, of the inner-cavity semi-major axis $\agap$ and eccentricity $\egap$ for the 2D Kepler-16 simulations.}
\label{fig:long_k16}
\end{figure}

\section{The physical models}
\subsection{The 3D model}
We solve the hydrodynamical equations for the conservation of mass,  momentum and internal energy in spherical coordinates $(r,\theta,\phi)$ (radial, polar, azimuthal), with the origin of the frame located at the centre of mass of the binary.  In particular, the energy equation that is incorporated in the code is given by:
 \begin{equation}
  \frac{\partial e}{\partial t}+{\bf \nabla}\cdot(e {\bf v})=-(\gamma-1) e {\bf \nabla}\cdot {\bf v}+\qcool ,
  \label{eq:energy}
 \end{equation}
where  ${\bf v}$ is the velocity,  and $e=\rho c_v T$ is the internal energy per unit volume, with  $\rho$  the density, $c_v$  the specific heat capacity at constant volume, and $T$ the temperature.  In the previous equation,  $\gamma$ is the adiabatic index, which is set to $\gamma=1.4$. 

We consider a cooling scheme where the temperature is relaxed towards a reference temperature $\tirr$ on a cooling timescale $\tcool$. In Eq.  \ref{eq:energy}, the  cooling rate ${\cal Q}_{\rm cool}$ is therefore given by:
\begin{equation}
\qcool=-\rho c_v\frac{T-\tirr}{\tcool},
\label{eq:qcool}
\end{equation}
Here, we assume $\tirr$ to correspond to the temperature set by  stellar irradiation. During the simulations, this temperature is regularly updated using the RADMC-3D Monte Carlo radiative transfer code (Dullemond et al. 2012), applying the procedure described below in Sect. \ref{sec:init}.  For the 3D models, the  temperature structure is updated every $100$ binary orbits. The  cooling timescale, $\tcool$,  is calculated every time step by considering 
the timescale for radiative loss of energy from a gaussian sphere with a characteristic length scale corresponding to the gas scale height $H$ (Lyra et al. 2016). It is given by:
\begin{equation}
\tcool=\frac{\rho c_v H \taueff}{3\sigma T^3},
\end{equation}
where $\sigma$ is the Stefan-Boltzmann constant, and where $\taueff$ is the effective optical depth given by:
\begin{equation}
\taueff=\frac{3}{8}\tau+\frac{\sqrt 3}{4}+\frac{1}{4\tau}.
\end{equation}
Following Bae et al. (2016), the optical depth $\tau$ is calculated as $1/\tau=1/\tauu+1/\taul$ with:
\begin{equation}
\tauu=\int_z^{z_{max}} \rho(z')\kappa(z')dz'
\end{equation}
and
\begin{equation}
\taul=\int_{z_{min}}^z \rho(z')\kappa(z')dz',
\end{equation}
where the opacity $\kappa$ is computed using the Rosseland mean opacity of Zhu et al. (2009).

\subsection{The 2D model}
In this work, we have also performed 2D simulations for direct comparison with the 3D models described above. These additional 2D simulations allow the circumbinary disc evolution to be  calculated over tens of thousands binary orbits for a reasonable computational cost, whereas evolving a 3D model beyond $\sim 2000$ binary orbits becomes prohibitive in terms of computational effort. 

In the 2D models, we consider the vertically integrated versions of the equations governing the disc evolution, and work in 2D $(R,\varphi)$ polar coordinates, with the origin of the frame located at the centre of mass of the binary. The energy equation that is solved is again given by Eq. \ref{eq:energy},  but now with the internal energy defined as $e=\Sigma c_v T$, with $\Sigma$ the surface density, and with the optical depth given by:
\begin{equation}
\tau=\frac{c_1}{\sqrt{2 \pi}}\kappa \Sigma. 
\end{equation}
In the previous equation, $c_1$ is a correction factor introduced to account for the drop of opacity with vertical height (Muller \& Kley 2011).   Again, we leave the description of the procedure employed to compute the irradiation temperature that appears in Eq. \ref{eq:qcool} to Sect. \ref{sec:init}. Compared to 3D calculations in which $\tirr$ are updated every $100$ binary orbits, we note that $\tirr$ is updated every $1000$ binary orbits in the 2D models.

\begin{table*}
\caption{Binary parameters employed in the simulations. Orbital parameters for the Kepler-16 system are taken from Doyle et al. (2011), and those for Kepler-34 are taken from Welsh et al. (2012). Stellar parameters have been taken from Georgarakos et al. (2021)}              
\label{table1}      
\centering                                      
\begin{tabular}{c c c c c c c c c}          
\hline\hline                        
System  &  $M_1(M_\odot)$ & $M_2(M_\odot)$ & $R_1(R_\odot)$ & $R_2(R_\odot)$ & $T_1(K)$ & $T_2(K)$ & $\abin$ (AU)  & $e_{\rm bin}$ \\
\hline
Kepler-16 & $0.69$ & $0.20$ & $0.65$ & $0.23$ & $4450$ & $3311$ & $0.22$ & $0.16$\\
Kepler-34 & $1.05$ & $1.02$ & $1.16$ & $1.09$ & $5913$ & $5867$ & $0.23$ & $0.54$\\
\hline                                             
\end{tabular}
\end{table*}

\begin{figure*}
\centering
\includegraphics[width=0.8\textwidth]{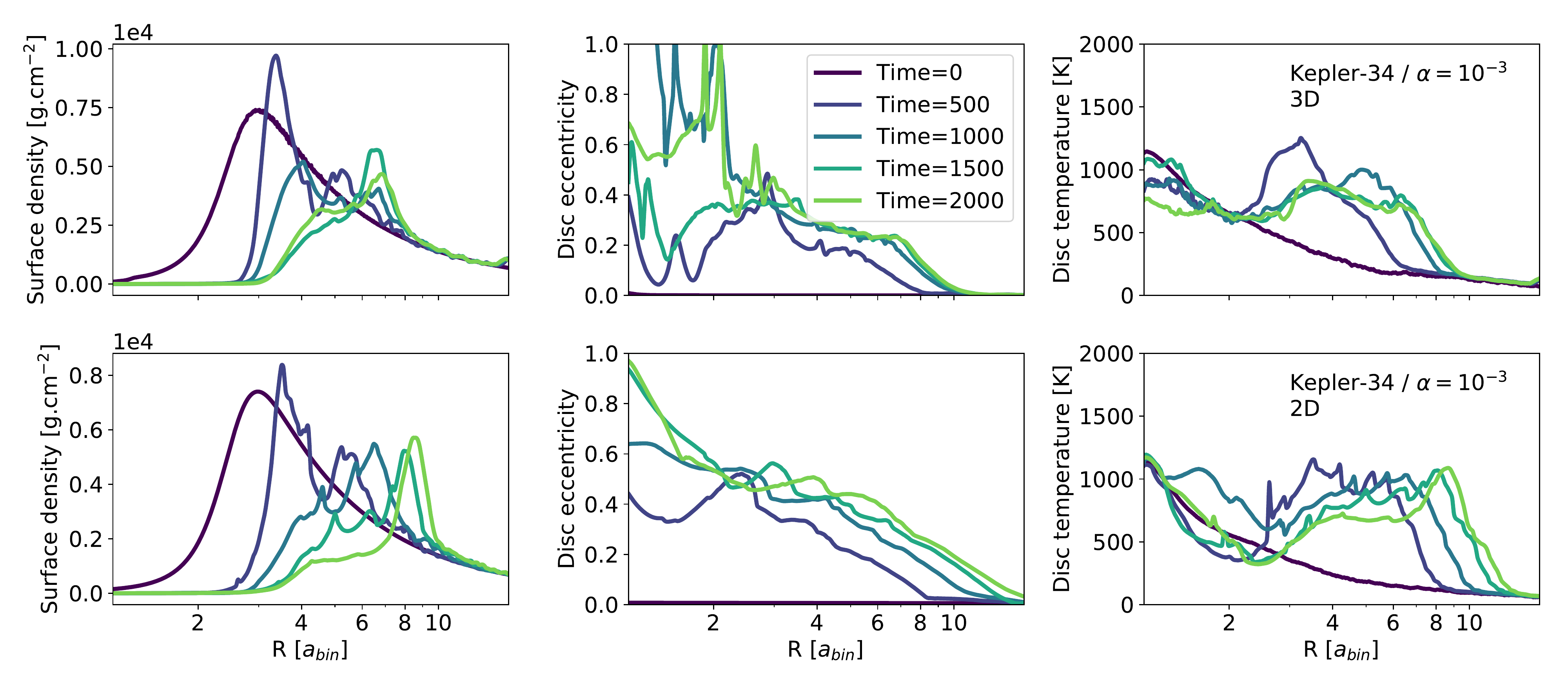}
\includegraphics[width=0.8\textwidth]{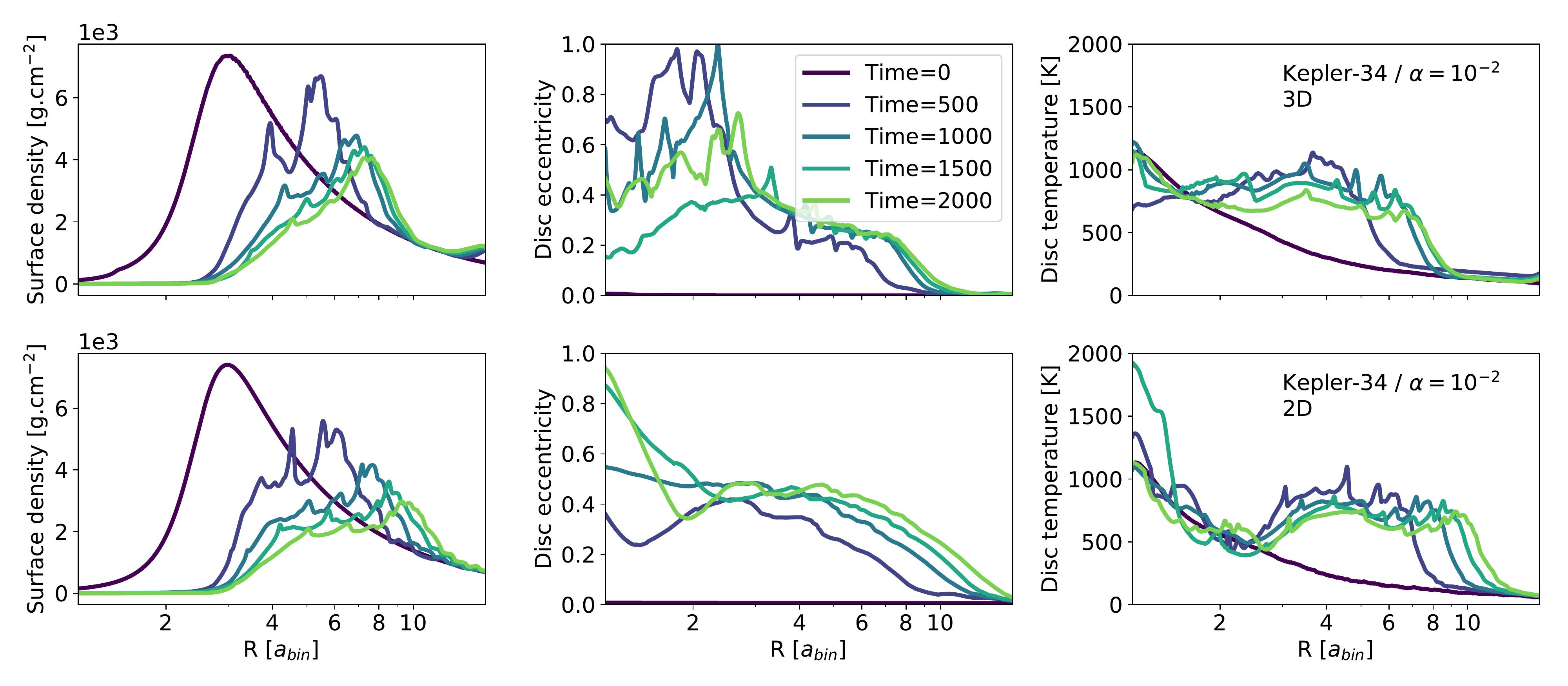}
\includegraphics[width=0.8\textwidth]{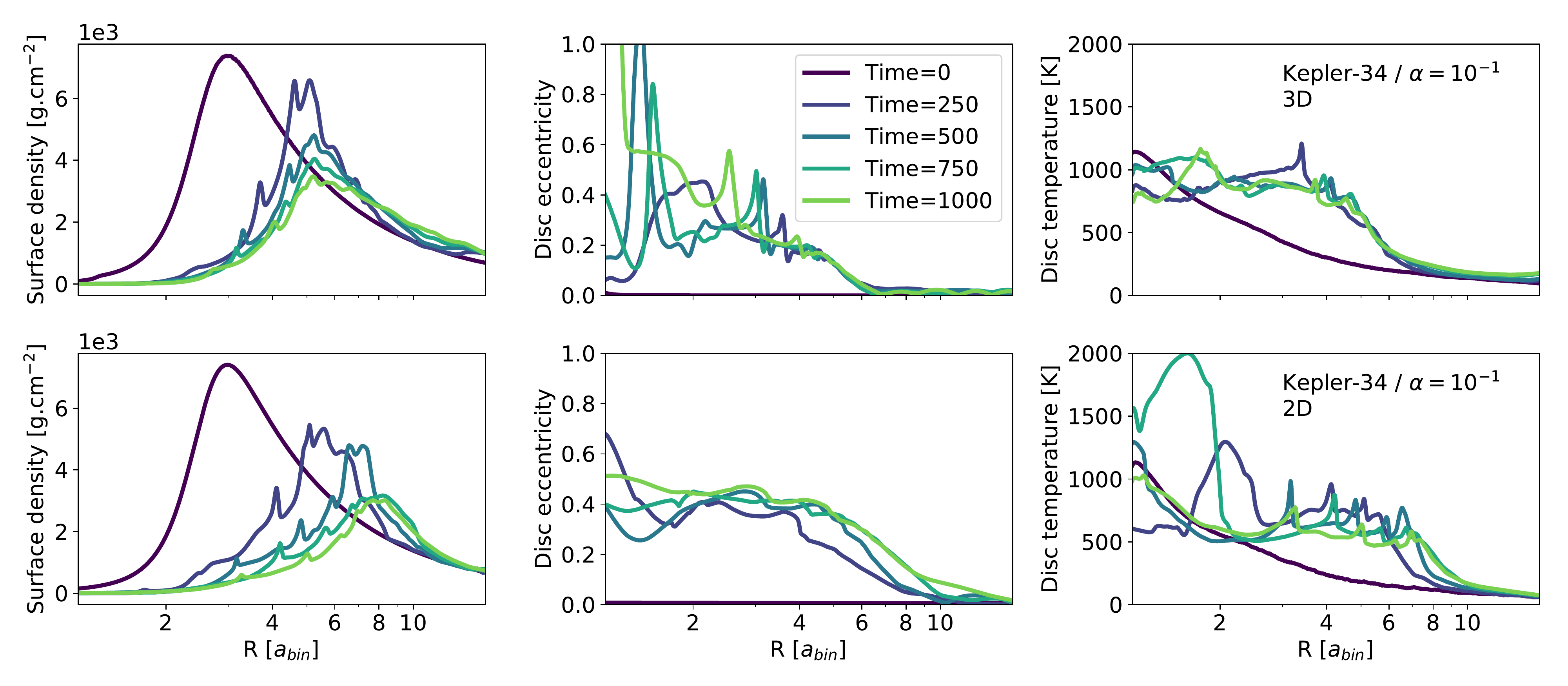}
\caption{Azimuthally averaged surface density (left panels), disc eccentricity (middle panels), and temperature (right panels) for the 2D and 3D models in the case of the Kepler-34 system.}
\label{fig:kep34_1}
\end{figure*}

\begin{figure*}
\centering
\includegraphics[width=\textwidth]{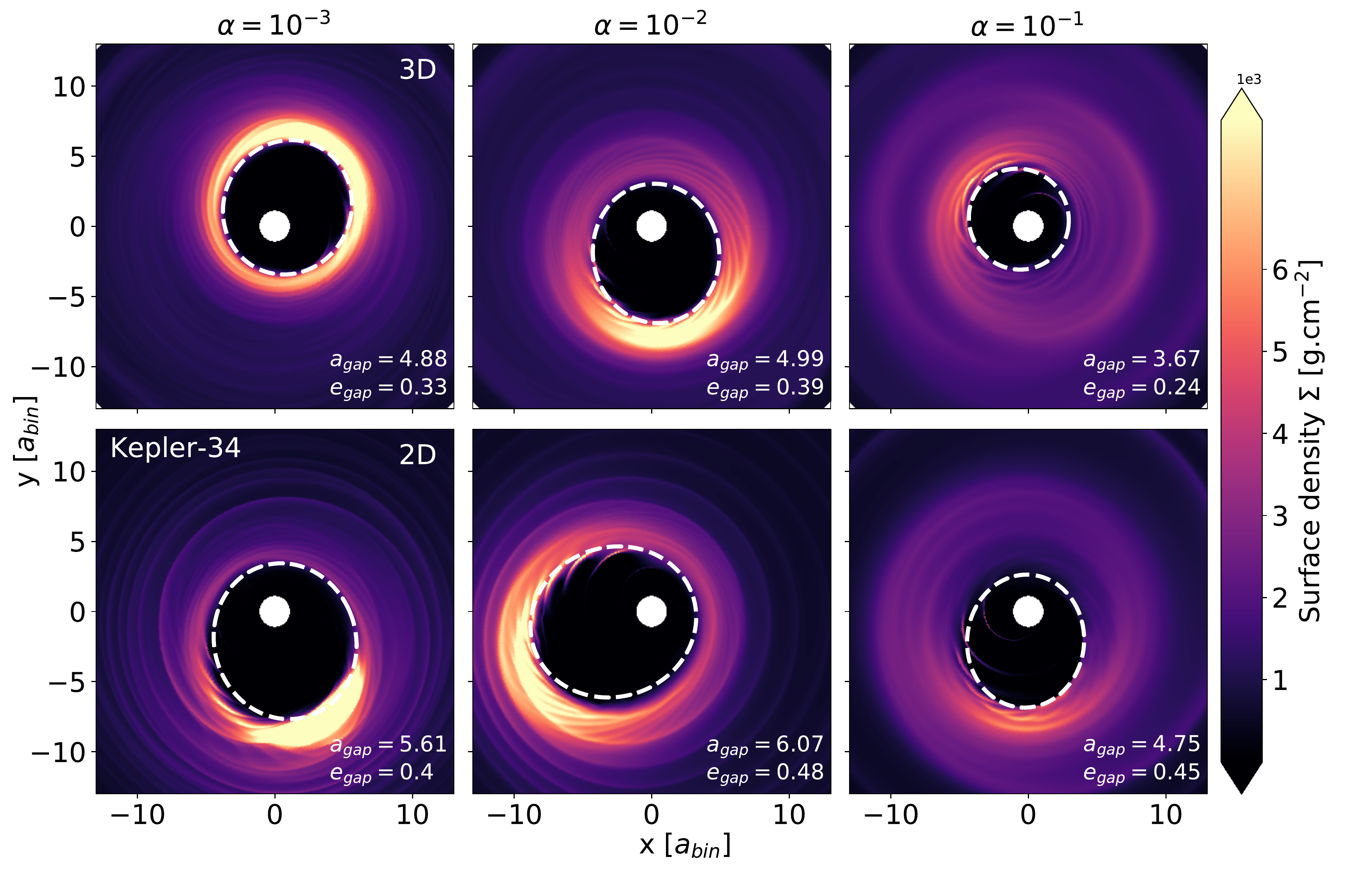}
\caption{For Kepler-34, 2D surface density distributions for the 2D and 3D models, and for different values of the $\alpha$ parameter. These snapshots correspond to a simulated time of  $2000$ $\Tbin$ for calculations   with $\alpha=10^{-3}-10^{-2}$, and of $1500$ $\Tbin$ for the run with $\alpha=0.1$.}
\label{fig:2d_kepler34}
\end{figure*}

\begin{figure}
\centering
\includegraphics[width=\columnwidth]{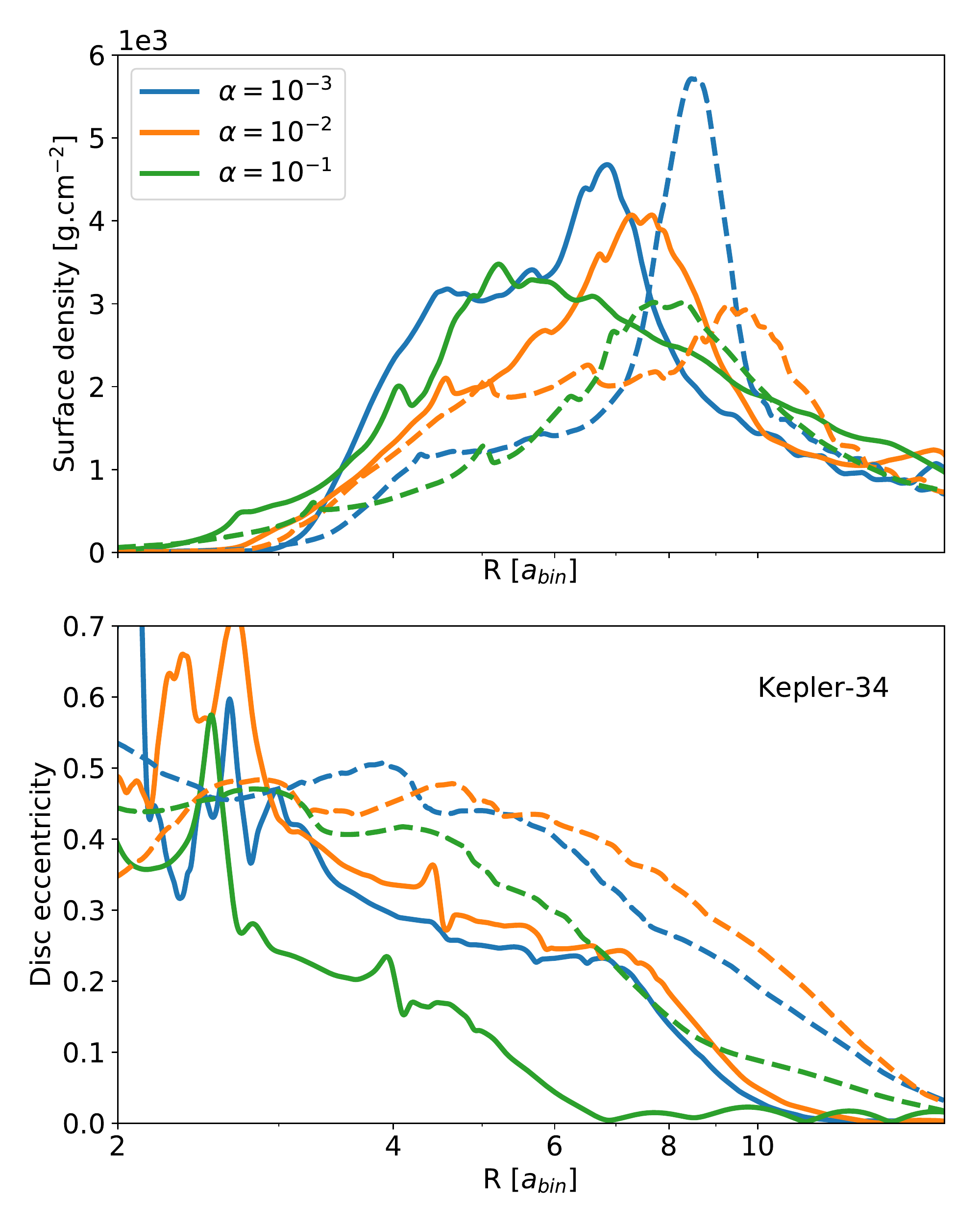}
\caption{For Kepler-34 and for each value of $\alpha$ we considered, comparison of the surface density (top panel) and eccentricity (bottom panel) profiles obtained in the 3D (solid lines) and 2D (dashed lines) models.}
\label{fig:profiles_k34_t1000}
\end{figure}

\begin{figure*}
\centering
\includegraphics[width=\textwidth]{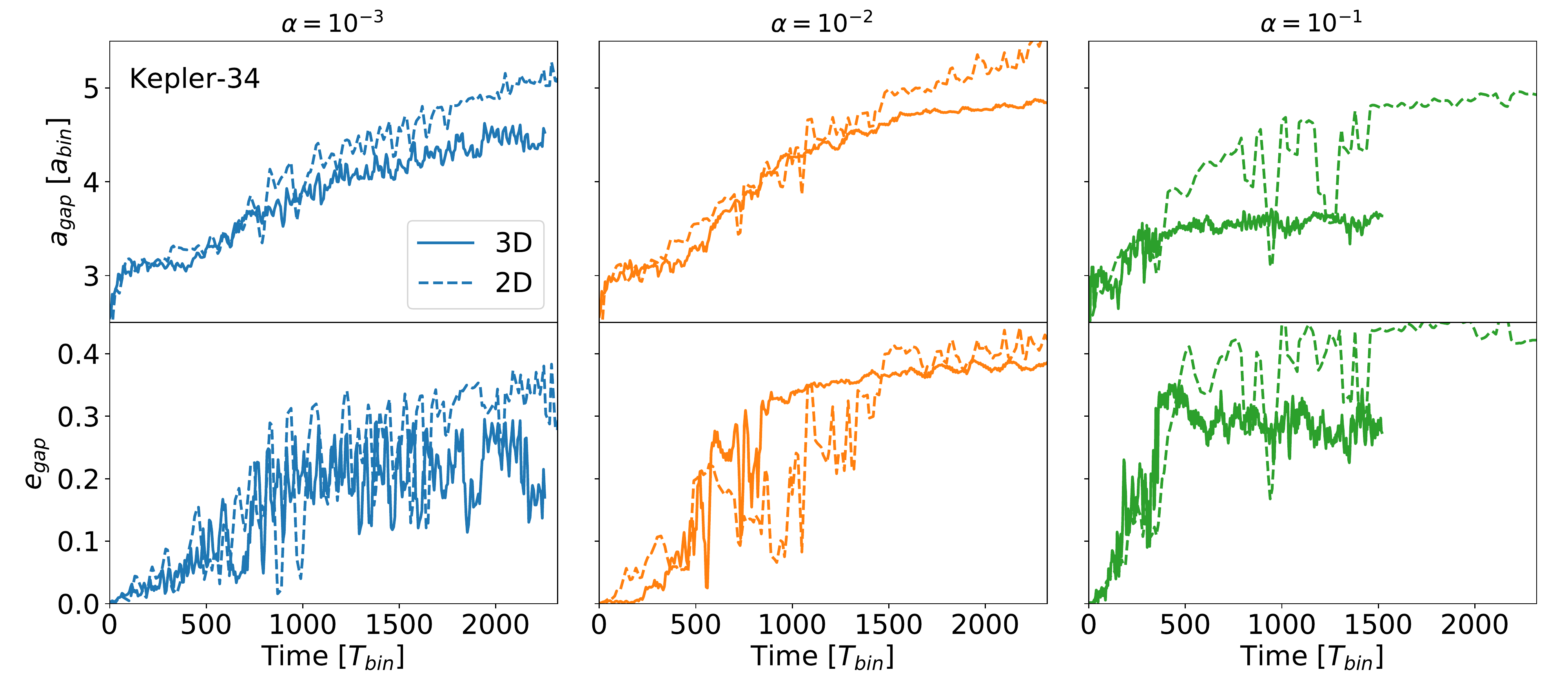}
\caption{For Kepler-34, time evolution of the inner cavity semi-major axis $\agap$ and eccentricity $\egap$ for each model.}
\label{fig:aegap_k34}
\end{figure*}

\begin{figure}
\centering
\includegraphics[width=\columnwidth]{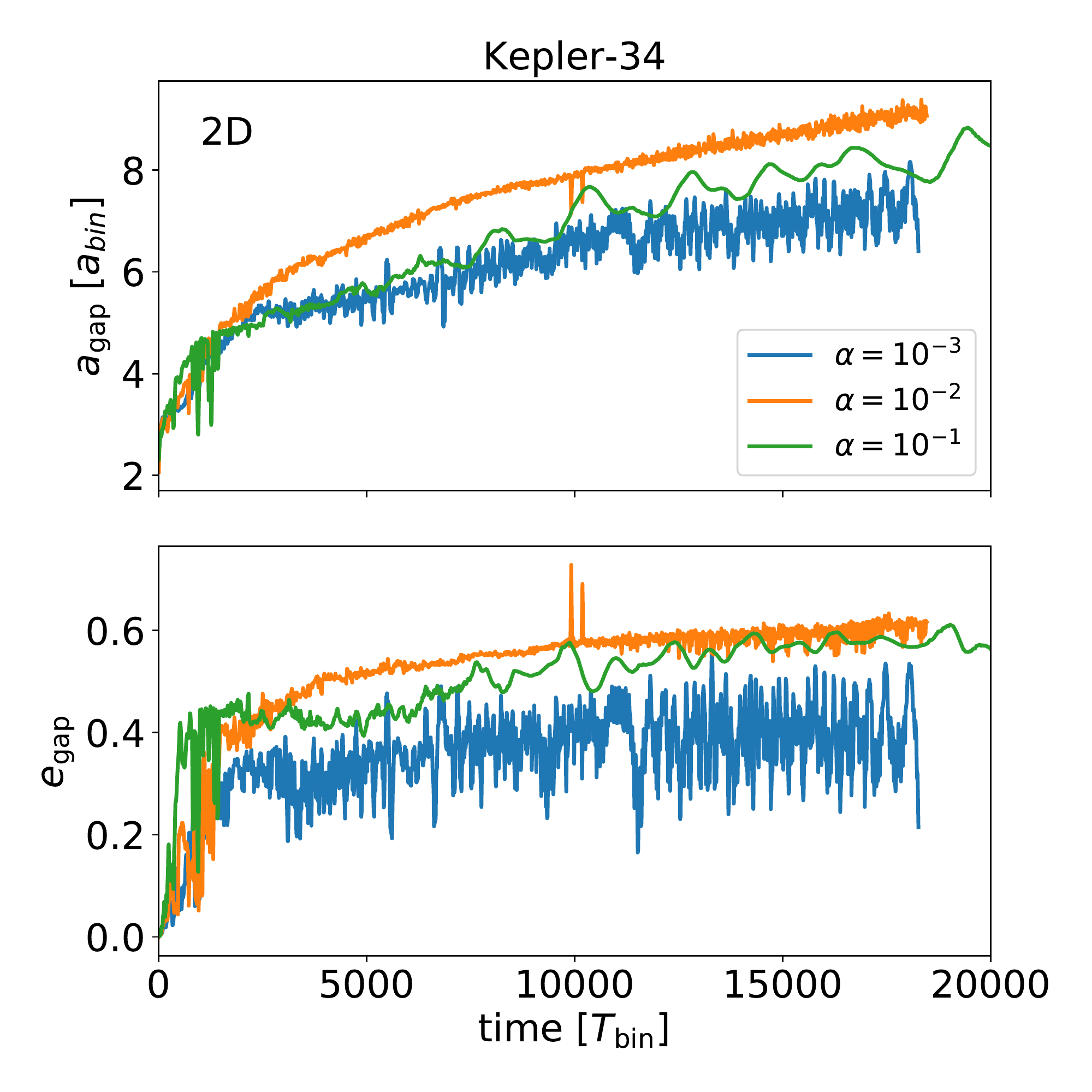}
\caption{For Kepler-34, time evolution over $2\times 10^4$ binary orbits of the inner cavity semi-major axis $\agap$ and eccentricity $\egap$ for each 2D model.}
\label{fig:long_k34}
\end{figure}

\section{The numerical model}
\subsection{Setup}
The simulations presented in this paper were performed using the multifluid version of FARGO3D (Benitez-Lamblay \& Masset 2016). 
Computational units are chosen such that the total mass of the binary is $M_\star=1$, the gravitational constant $G=1$, and the radius $R=1$ in the computational domain corresponds to the binary semi-major axis $\abin$. When presenting the simulation results, unless otherwise stated we use the binary orbital period $\Tbin=2\pi\sqrt{\abin^3/GM_\star}$ as the unit of time.

 In the three-dimensional models, the computational domain in the radial direction extends from $R_{\rm in}=1.13\;\abin$ to $R_{\rm out}=18\;\abin$ and we employ $995$ logarithmically spaced grid cells.  Ideally,  the inner boundary should be set to $R_{\rm in}\approx \abin$ to allow for accurate modelling of mass flow and angular momentum transfer around the binary (Mutter et al. 2017).  As shown by Thun \& Kley (2017), however, setting $R_{\rm in}=1.13\;\abin$ appears to be a good compromise between an increased run time  due to a smaller timestep and realistic modelling.  For $R_{\rm in}\lesssim1.13\;\abin$, the change in surface density appears to be very small upon further reduction of  $R_{\rm in}$.  In the azimuthal direction the simulation domain extends from $0$ to $2\pi$ with $700$ uniformly spaced grid cells. In the meridional direction, the simulation domain covers $3.5$ disc pressure scale heights above and below the disc midplane, and we adopt $144$ uniformly spaced grid cells. 
 
 In the two-dimensional models, the computational domain in the radial direction extends from $R_{\rm in}=1.13\;\abin$ to $R_{\rm out}=40\;\abin$ and we employ $1280$ logarithmically spaced grid cells, such that the radial resolution is equivalent to that used in the 3D models. Compared to 3D models, the location of the outer boundary in the 2D simulations is set to a larger value due  to a clear tendency for getting larger inner cavities when using a 2D setup. Numerical experiments have demonstrated that the location of the outer radius has no effect on the global evolution  provided that surface density at the outer boundary is not impacted by the formation and evolution of the inner cavity, namely provided it remains close to its initial value.  In the azimuthal direction, the resolution is also set to be similar to that used in 3D.

\subsection{Initial conditions}
\label{sec:init}

The initial surface density profile in the 2D and 3D models is given by:
\begin{equation}
\Sigma=f_{\rm gap}\, \Sigma_0\left(\frac{R}{R_0}\right)^{-3/2}
\label{eq:sigma0}
\end{equation}
where  $\Sigma_0$ is the surface density at $R_0=\abin$ and which is set to $\Sigma_0=1.3 \times 10^{-4}$ in code units. $f_{\rm gap}$ is a gap-function used to initiate the disc with an inner cavity (assumed to be created by the binary), and is given by:
\begin{equation}
f_{\rm gap}=\left(1+\exp\left[-\frac{R-R_{\rm gap}}{0.1R_{\rm gap}}\right]\right)^{-1},
\end{equation}
 where $R_{\rm gap}=2.5\abin$ is the analytically estimated gap size (Artymowicz \& Lubow 1994). The initial azimuthal velocity is set to the Keplerian velocity, whereas the initial radial velocity is set to zero. In the 3D runs,  the meridional velocity is also set to zero.

The initial temperature $T_0$ is determined by performing a Monte Carlo Radiative Transfer calculation of photon propagation using  RADMC-3D, in which multiple radiation sources can be incorporated. Here, we consider two radiation sources whose stellar parameters correspond to the  Kepler-16 (AB) or Kepler-34 (AB) systems. The stellar parameters that we employed for these two systems can be found in Table 1. 

In  RADMC-3D, we assume a grain size distribution that ranges from $0.1$ $\mu m $ to $10$ $\mu m $ with power-law exponent of $-3.5$. The grains are composed of 60 \% silicate and 40 \% amorphous carbon, with an internal density of 2.7 $\rm g \; cm ^{-3}$. The optical constants of silicate and amorphous carbon are obtained from Draine \& Lee (1984) and Li \& Greenberg (1997). For the 3D models,  we first use the initial three dimensional gas density distribution to determine dust densities,  assuming a net dust-to-gas ratio of $1\%$. RADMC-3D  then computes the dust temperature  and since $\mu$m-sized  grains tend to be strongly coupled to the gas, we can safely consider the gas and dust temperatures to be equal.   $T_0$ can  therefore be simply obtained by performing an azimuthal average of the calculated gas temperature. We illustrate this procedure by showing in the upper panel of Fig.~\ref{fig:temp0_k34} the resulting temperature structure for the Kepler-34 analogue simulation calculated by RADMC-3D. The bottom panel of  Fig.~\ref{fig:temp0_k34} displays the two dimensional (R-Z) distribution of the dimensionless cooling time $\beta=\tcool \Omega_k$. At the location of the inner cavity,  we see that $\beta \approx 100$ so that the disc behaves almost adiabatically, whereas  it is close to being isothermal near the disc surfaces. For the 2D models, we first assume vertical hydrostatic equilibrium to deduce the three dimensional gas density distribution.  We then apply a  method similar to that described in the context of 3D models  to get the initial gas temperature, but with the exception that in 2D it is defined as the vertically integrated temperature distribution (e.g. Bae et al. (2019): 
 \begin{equation}
 T(R)=\frac{\int \rho(R,Z) T(R,Z) dZ}{\int \rho(R,Z) dZ}
 \end{equation}
  From this previous relation,  the initial disc aspect ratio can be computed and is given by:
  \begin{equation}
 h=\left(\frac{{\cal R} T(R)}{\mu}\right)^{1/2}v_k^{-1}
 \end{equation}
 where ${\cal R}$ is the perfect gas constant,  $\mu$ is the mean molecular weight which is set to $\mu=2.4$,  and $v_k$ the Keplerian velocity. The initial surface density, temperature, and aspect ratio are displayed in Fig.~\ref{fig:disc0}. The factor of $\sim 2$ difference that is observed between the initial temperatures at the inner edge of the disc is in line with expectation. The received energy per unit time at this location is given by (Bitsch et al. 2013):
 \begin{equation}
 Q_+=2 \pi \rin 2H F_\star
 \end{equation}
 where $F_\star$ is the stellar flux which can be approximated by:
  \begin{equation}
F_\star=\sigma \frac{R_1^2T_1^4+R_2^2T_2^4}{\rin^2},
 \end{equation}
 where $T_1,T_2$ ($R_1,R_2$) are the temperatures (radii) of the primary and secondary stars respectively. On the other hand, the cooling at the surface of the inner disc edge is given by:
  \begin{equation}
\qcool=-4 \pi \rin H \sigma T_D^4,
 \end{equation}
 where $T_D$ is the disc temperature. In thermodynamical equilibrium, we therefore have:
   \begin{equation}
T_D=\left(\frac{R_1^2T_1^4+R_2^2T_2^4}{\rin^2}\right)^{1/4}.
 \end{equation}
 From the latter equation, and using the stellar parameters given in Table 1,  it is straightforward to recover the factor $2$ shift between the disc temperatures of the Kepler-16 and Kepler-34 systems that are observed in Fig. \ref{fig:disc0}.
 
 For Kepler-34,    the temperature at the inner edge of the disc is likely to be higher than the threshold temperature for magnetorotational turbulence through thermal ionization, which is estimated to $T_{MRI}\approx 800-1200$ K (e.g. Flock et al. 2016). For Kepler-16, parametric instabilities may also give rise to turbulence, with an $\alpha$ viscous stress parameter (Shakura \& Sunyaev 1973) $\alpha \approx 5\times 10^{-3}$ (Pierens et al. 2020).  Regardless of the origin of turbulence, the anomalous viscosity in the disc is modelled using the standard alpha prescription for the effective kinematic viscosity $\nu =\alpha c_s H$, where in this work, we use values of $\alpha = 10^{-3}, 10^{-2}$ and $10^{-1}$. Although a value of $\alpha=10^{-1}$ is  probably too large  for  a standard protoplanetary disc, it is comparable to the  value expected in circumbinary disks around binary black holes (Moody et al. 2019; Munoz et al. 2019;  Tiede et al. 2020). Here, the  calculation with $\alpha=0.1$ serves for us as a benchmark simulation used to validate the results from other runs, since it is expected to converge more rapidly due to a shorter viscous timescale. It is important to note that we do not incorporate in the energy equation the term corresponding to viscous heating so that the vertical thermal structure is similar for a given set of binary parameters. Our aim here is rather to examine the effect of viscosity on angular momentum transport rather than on the disc thermodynamics. Nevertheless, we discuss the effect of including viscous heating in Appendix A. 

\subsection{Boundary conditions}
We adopt an outflow boundary condition at the inner edge to allow mass to accrete onto the binary. For an outflow boundary condition, the velocity is set to 0 if directing toward the disc,  such that gas material can leave the computational grid but no mass can flow into the domain. The values for the gas density and internal energy in the ghost zones have the same values as in the first active zone. In the context of 2D models, it has been shown that  employing an open boundary together with a location of the inner disc edge $\rin \approx \abin$ leads to a quasi-stationary disc structure after ${\cal O}(10^4)$ binary orbits. Compared to other boundary conditions, adopting an open boundary also appears  to be less sensitive to numerical issues (Thun \& Kley 2017). 

At the outer radial boundary, we employ a wave-killing zone for $R > 0.88$ $\rout$  to avoid wave reflection/excitation at the disc outer edge (de Val-Borro et al. 2006). The impact of this wave-killing zone on the disc shape and eccentricity is expected to be small since, as we have mentioned earlier, the disc structure near the wave-killing zone remains close to initial one.

At the meridional boundaries of the three dimensional domain, an outflow boundary condition is also employed, but with the exception that for the  gas density we follow Bae et al. (2016a) and maintain vertical stratification by solving the following condition for hydrostatic equilibrium:
\begin{equation}
\frac{1}{\rho}\frac{\partial}{\partial \theta} (\cs^2 \rho)=\frac{v_\phi^2}{\tan \theta}.
\end{equation}

\section{Results}
\label{sec:results}

\subsection{Kepler-16}
In this section, we describe the results for the Kepler-16 analogue models, the parameters for which are given in Table 1.

\subsubsection{Comparison between 3D and 2D models}
In Fig.~\ref{fig:kep16_1}, we compare the surface density, disc eccentricity and temperature profiles for the 2D and 3D models . The computational expense of running 3D simulations does not allow simulations to run for evolution times $\gtrsim 2000$ $\Tbin$, so we restrict the comparison to times $t \le 2000$ $\Tbin$.  For the 3 dimensional runs, the disc eccentricity $e_d(R)$ at radius $R$ is defined as a density-weighted average,  obtained by averaging over $\theta$ and $\varphi$ the quantity $e_c$, where $e_c$  is the eccentricity of a fluid element computed at the center of each grid cell:
\begin{equation}
e_d(R)=\frac{\int_{\theta_-}^{\theta_+}\int_0^{2\pi} \int_R^{R+dR} \rho e_c dV }{\int_{\theta_-}^{\theta_+}\int_0^{2\pi} \int_R^{R+dR} \rho  dV}
\end{equation}
where $dV$  is the volume of one grid cell and where the integral over $\theta$ is performed over the total vertical extent of the disc. For the two dimensional models, $e_d(R)$ is rather calculated following Pierens \& Nelson (2013) and  is given by:

\begin{equation}
e_d(R)=\frac{\int_0^{2\pi} \int_R^{R+dR} \Sigma e_c dS }{\int_0^{2\pi} \int_R^{R+dR} \Sigma  dS}
\end{equation}
 where  $dS$ the surface area of one grid cell.
 
Consistent with previous work (Pierens \& Nelson 2013; Kley et al. 2019; Penzlin et al. 2021), the 2D models show  larger and more eccentric inner cavities as $\alpha$ is increased.  Interestingly, we find the opposite behaviour in the 3D models. Although the location of the maximum in the surface density seems similar for  $\alpha=10^{-3}$ and $\alpha=10^{-2}$ in the 3D models, comparing the disc eccentricity panels reveals a higher disc eccentricity when $\alpha=10^{-3}$. Increasing $\alpha$ up to $\alpha=0.1$ leads to a  smaller cavity size, while growth of the disc eccentricity is not observed at all in this case.  

Due to  higher cavity eccentricities, the 2D models tend to lose more mass through the inner boundary,  since gas is more prone to approach the inner edge of the disc in that case.  This is illustrated  in Fig.~\ref{fig:discmass_k16} where we plot the evolution of the disc mass enclosed within the computational domain. Here, the 3D simulations are  represented by solid lines whereas the dashed lines correspond to 2D models.  The  increase of the disc mass observed in the 3D simulation with $\alpha=0.1$ is a consequence of i) the wave-killing zone employed near the outer boundary, which acts as a mass reservoir because at each radius the radial velocity in the disc is towards the star; plus ii) the fact there is no growth of the disc eccentricity for this model such that inner disc material to does not approach close to the inner boundary . A similar effect is not found in 2D due to the higher value for $\rout$, resulting in a longer viscous timescale at the outer boundary.

In Fig. \ref{fig:2d_kepler16} we present images of the surface density distributions for the different models, after an evolution time of $2000$ $\Tbin$ when $\alpha=10^{-3}$ and $10^{-2}$, and  $1500 \Tbin$ when $\alpha=0.1$. The over-plotted dashed  white line 
traces the ellipse that best fits the structure of the inner cavity.  The parameters of the ellipse are calculated following Thun et al. (2017), who assume that the focus of the ellipse is located at the centre of mass of the binary and the apocentre is in the direction where the surface density has its maximum value, and at the location where the surface density is 10\% of the maximum value.  The values reported in the figure for the cavity semi-major axis, $\agap$, and eccentricity, $\egap$, show that the cavity structures in 2D and 3D are very similar when $\alpha=10^{-3}$, and also confirm that increasing $\alpha$ produces smaller cavities and disc eccentricities in 3D, whereas in 2D it produces larger cavity sizes and a eccentricities (at least after a run time of $2000 \Tbin$). One implication of this trend is that for $\alpha \ge 0.01$, not only the size but also the eccentricity of the inner cavity are smaller in 3D than in 2D, with the important consequence that migrating circumbinary planets should park closer to the binary when adopting a 3D setup and these higher values of $\alpha$.

The time evolution of $\agap$ and $\egap$ over $2000$ binary orbits is plotted for each model in Fig.~\ref{fig:aegap_k16}. Both quantities show significant fluctuations because the procedure of finding the apocentre of the cavity may be misguided by local density maxima, as noticed by Penzlin et al. (2021). Regarding the long term trends, however, it is expected that the time to reach equilibrium should be shorter when $\alpha=0.1$ compared to the lower values, and it does indeed seem to be the case that the corresponding 3D run has converged after $\sim 10^3$ $\Tbin$. At that time, both $\agap$ and $\egap$ appear to have reached saturated values. For lower values of $\alpha$, however, it is difficult to assess whether or not the 3D calculations have really converged after the total simulated time of $2000$ orbits.  For $\alpha=10^{-3}$ it appears that continuing evolution of $\agap$ and $\egap$ is occurring, whereas for $\alpha=10^{-2}$ the values appear to be more steady on average towards the end of the run time. In spite of the ambiguity about a steady state being achieved in the 3D runs, it is clear that there is a general tendency for the values of $\agap$ and $\egap$ obtained in the 3D runs to trend below the values obtained in 2D, and for $\alpha \ge 10^{-2}$ the cavities obtained in 3D are significantly smaller and more circular compared to those obtained in the 2D simulations.  This can also be seen in Fig. \ref{fig:profiles_k16_t1000} where we compare for each value of $\alpha$,  the disc surface density and eccentricity profiles obtained at the end of the 3D runs with the corresponding 2D ones obtained at the same time.

\subsubsection{Long timescale evolution of the 2D runs}
Previous work has shown that in the context of a 2D setup, circumbinary discs typically settle to a stationary state after ${\cal O}(10^4)$ binary orbits.  ${\cal O}(10^5)$ binary orbits can even be required to reach equilibrium in radiative discs, as shown recently by Kley et al. (2019) and  Sudarshan et al. (2022). Our thermodynamical treatment of the disc is different in comparison to these previous studies in this work because the main source of heating is stellar irradiation rather than viscous heating. However, it is clear that a total evolution time of $2000$ $\Tbin$ is not enough to allow the 2D discs to settle to a stationary state.  For this set of runs, the time evolution of $(\agap,\egap)$  over a longer simulated timescale of $2\times 10^4$ binary orbits is presented in Fig. \ref{fig:long_k16}. Not surprisingly, the model with $\alpha=0.1$ reaches a quasi-stationary state earlier compared to the two other runs, for which both $\agap$ and $\egap$ are still evolving  at $2\times 10^4$ binary orbits.  We note, however, that in spite of reaching a quasi-steady state, the 2D run with $\alpha=0.1$ shows strong fluctuations in the computed values of $a_{\rm gap}$ and $e_{\rm gap}$. Inspection of the results indicates that this arises because of the way that the algorithm used to calculate these quantities first identifies the location of the grid cell containing the maximum density. This is found to fluctuate strongly in this run, possibly because the cavity is relatively small and has lower eccentricity. Hence, the fluctuations are a consequence of the algorithm somewhat misbehaving in this case.  Since our primary intention here is to focus on 3D models, we did not explore the subsequent circumbinary disc evolution until a quasi-stationary state is reached for these 2D runs. Nevertheless, we can make use of the results of past studies on radiative circumbinary disc evolution to guess the evolution outcome in our 2D models with $\alpha=10^{-3}$ and $10^{-2}$.  The case $\alpha=10^{-3}$ has been examined by Sudarshan et al. (2022) using a $\beta$-cooling prescription. For $\beta \approx 100$, which as mentioned in Sect. \ref{sec:init} corresponds to the value for $\beta$ in the midplane of our disc models, these authors found $\agap=4.35$ $\abin$  and $\egap=0.28$ at equilibrium. We see that despite a slightly different treatment of the thermal physics, we obtain values for the cavity semi-major axis and eccentricity which are in rather good agreement with the ones reported by Sudarshan et al. (2022).  The case with $\alpha=10^{-2}$  has been considered by Kley et al. (2019), but only for a binary eccentricity $\ebin=0.1$. For this value of $\ebin$, they found $\agap\approx 4 \; \abin$ and  $\egap \approx 0.38$ once a quasi-steady state is reached. Although we obtain a disc eccentricity similar to that found by Kley et al. (2019), we find a slightly higher value for the cavity semi-major axis $\agap\approx 5$. This discrepancy probably arises because of the different values for $\ebin$ that have been used, and/or possibly because  we  omit viscous heating in our simulations.

\begin{figure}
\centering
\includegraphics[width=\columnwidth]{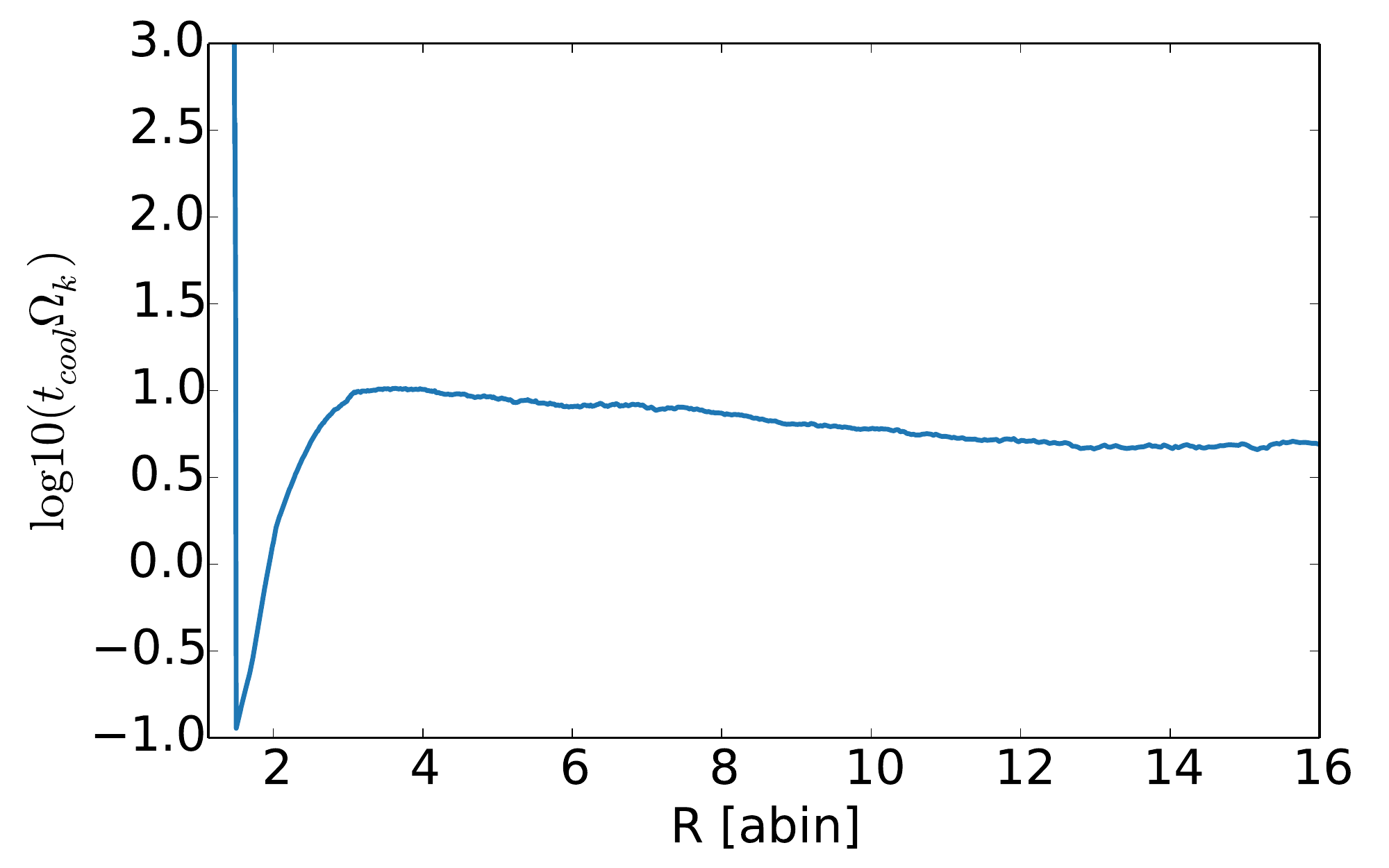}
\caption{Radial profile of the  dimensionless cooling time for the initial disc models. }
\label{fig:tcool_vs_r}
\end{figure}

\begin{figure*}
\centering
\includegraphics[width=\textwidth]{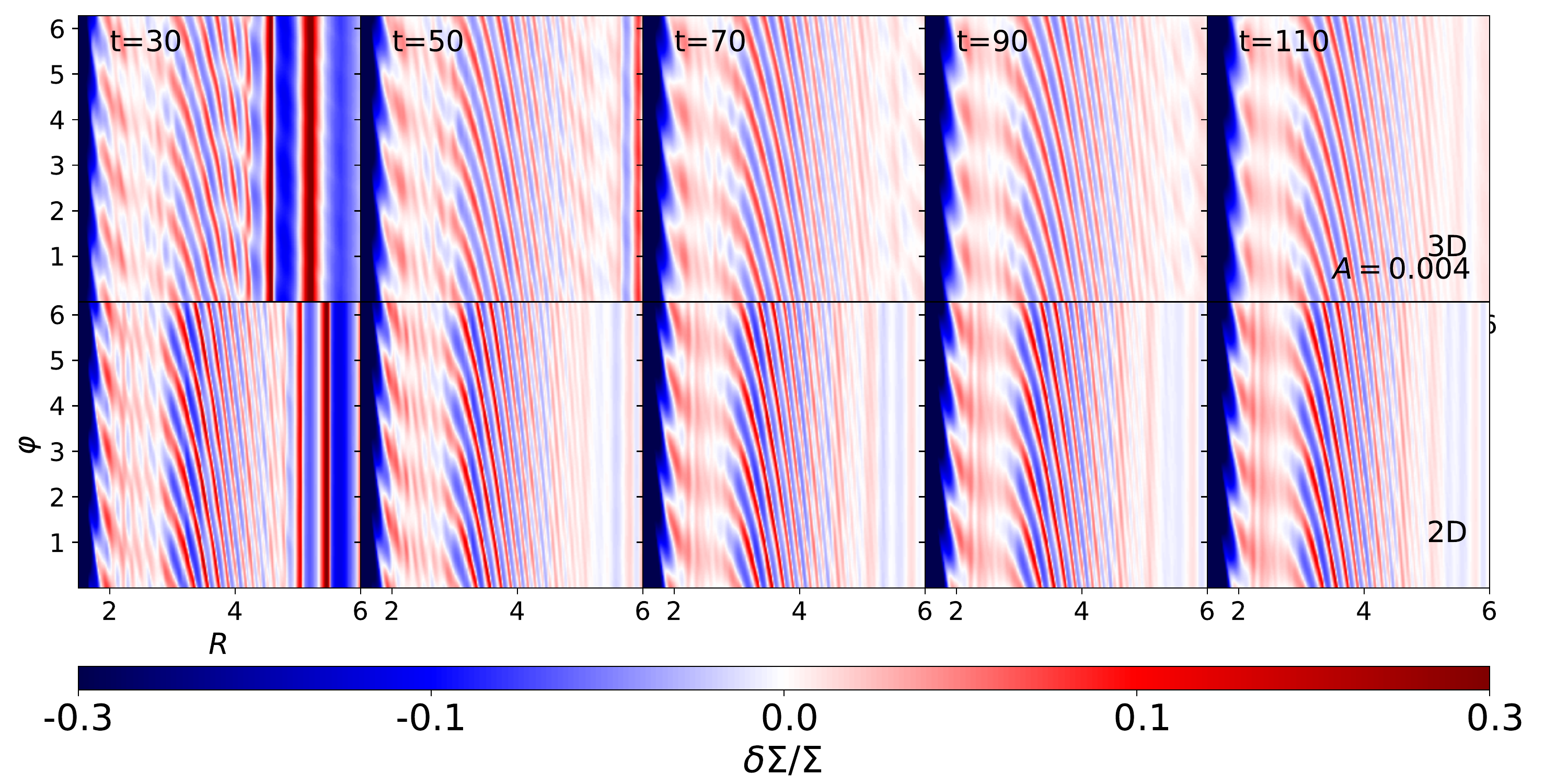}
\includegraphics[width=\textwidth]{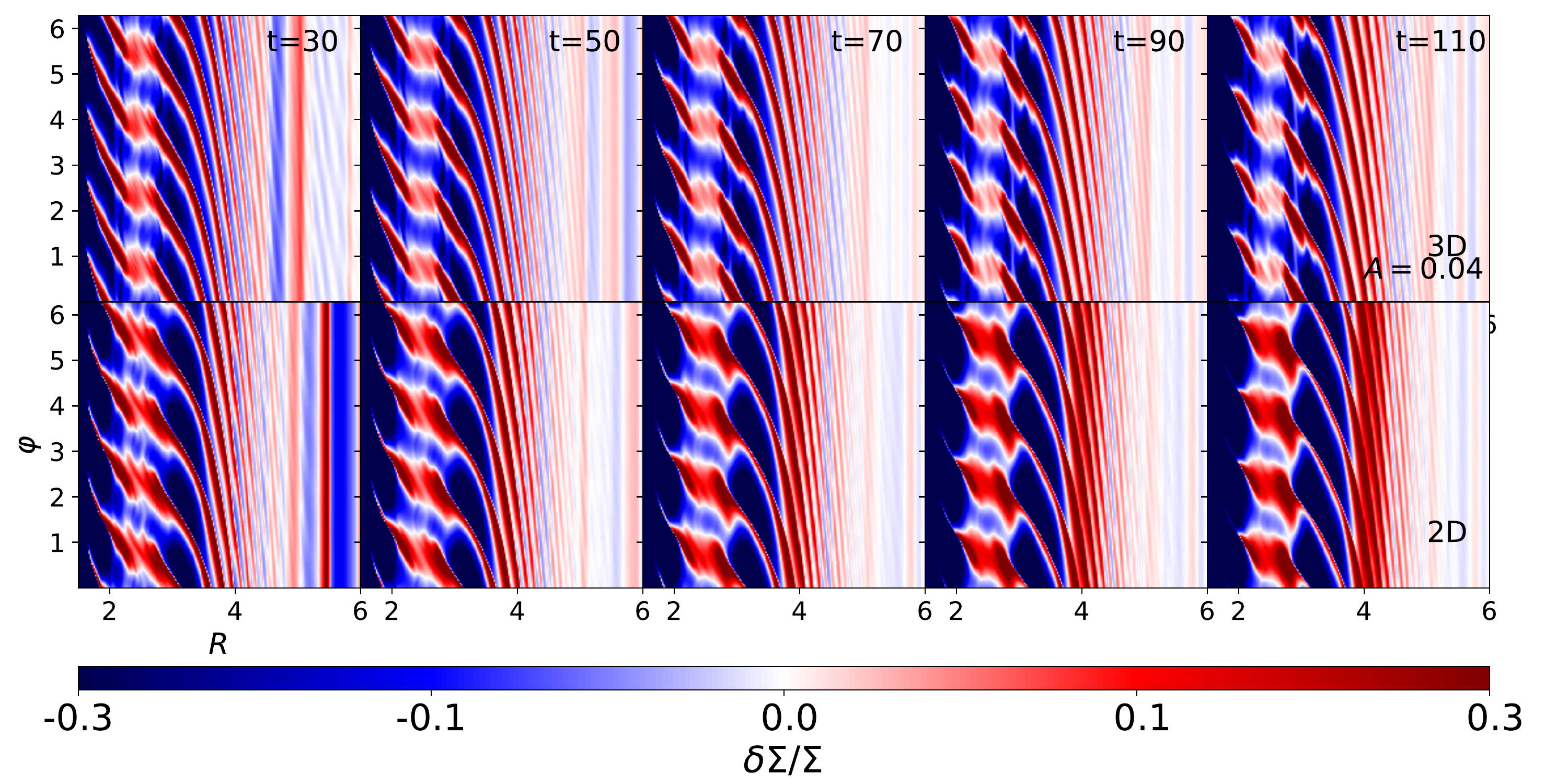}
\caption{Top panel: Surface density perturbations in the 3D and 2D discs using the perturbing potential given in Eq.~\ref{eq:pert_pot} with $A=0.004$. Bottom panel: same but for a perturbation amplitude $A=0.04$.}
\label{fig:pert_pot}
\end{figure*}

\subsection{Kepler-34}
We now turn our attention to the circumbinary disc evolution for the Kepler-34 analogue system (see Table 1 for the binary parameters). In Fig~\ref{fig:kep34_1}, we show the surface density, disc eccentricity and temperature profiles for the 2D and 3D models at times $\le 2000$ $\Tbin$. Contrary to Kepler-16 simulations, growth of disc eccentricity is now observed in the 3D run with $\alpha=0.1$, although the disc eccentricity is smaller by a factor of $\sim 2$ in comparison with its 2D counterpart. This tendency of the disc eccentricity being smaller in 3D is also verified in the simulations with lower values of $\alpha$. Hence, as with the Kepler-16 analogues, we find that the the cavity sizes are systematically larger and more eccentric in 2D than in 3D. Looking at the two-dimensional density distributions and the elliptical fits to the shape of the inner cavity in Fig.~\ref{fig:2d_kepler34} confirms this tendency.  Again, this would imply that migrating protoplanets should park closer to the central binary when adopting a 3D setup, but this needs to be checked in the future using dedicated simulations of embedded planets in 3D circumbinary disc models. Although we expect planets  around highly eccentric binaries to stop their migration slightly beyond the location corresponding to the cavity semi-major axis (Penzlin et al. 2021), it is interesting to notice that for $\alpha \lesssim 0.01$, the values obtained for $\agap$ in the 3D models are close to the observed semi-major axis of Kepler-34b which is $a_p\approx 4.7\abin$.

With the exception of the run with $\alpha=0.1$, the values reported for the ellipse parameters in Fig. \ref{fig:2d_kepler34} suggest that $\agap$ and $\egap$ increase  with $\alpha$ in both 3D and 2D discs,  contrary to what has been obtained in the Kepler-16 simulations. This similar behaviour between the 2D and 3D models can also be seen in Fig.~\ref{fig:aegap_k34} where we plot $\agap$ and $\egap$ as a function of time. Compared to the Kepler-16 simulations, here it appears more evident that the 3D models reach a quasi-stationary state earlier than their 2D counterparts. While it is not possible to verify beyond all doubt that the 3D disc models have reached a quasi-steady state for the runs with $\alpha<10^{-1}$, it is clear that for all values of $\alpha$ the cavity semi-major axes and eccentricities obtained in the 3D runs trend significantly below the values obtained in the corresponding 2D runs. Hence, the conclusion that the cavity sizes and eccentricities obtained in 3D are smaller than those obtained in 2D appears to be robust.  For Kepler-34, we compare in  Fig. \ref{fig:profiles_k34_t1000}  the disc surface density and eccentricity profiles obtained at the end of the 3D runs with the 2D profiles at the same time. The fact that the cavity sizes and disc eccentricities tend to be smaller in 3D is here clearly visible.  

 For the 2D runs, the long-term evolution of the cavity semi-major axis and eccentricity is presented in Fig. \ref{fig:long_k34}. Similar to the discs around the Kepler-16 analogues, an equilibrium state has not been reached after an evolution time of $2\times 10^4$ $\Tbin$. At that time, the cavity size is estimated to $\agap\sim 8\abin$ for $\alpha=10^{-2}$, whereas we find $\agap \sim 7 \abin$ for $\alpha=10^{-3}$. This latter value is slightly higher that the one reported by Sudarshan et al. (2022), who found $\agap \sim 6 \abin$ for similar parameters in their simulations using a  $\beta$ prescription for cooling. This discrepancy possibly arises because we omit viscous heating, resulting in a smaller viscosity $\nu=\alpha c_s H$ because of the lower temperature. As a consequence, the viscous torques that tend to  counteract the gravitational torques from the binary are smaller in our simulations, leading to larger cavity sizes.
 
 Overall, our results show that there is trend for 3D models to reach equilibrium earlier than 2D models, and for the cavity sizes and eccentricities in 3D models to be smaller than in their 2D counterparts. 

\section{Discussion}
In this section, we investigate the origin of the differences that are observed between the 2D and 3D models.  We begin the discussion by emphasizing that the spiral waves excited by the binary carry angular momentum that will be deposited in the disc to induce the formation of the inner cavity. Here,  the larger cavity sizes obtained in the 2D simulations suggest that the angular momentum flux (AMF) carried by the spiral waves is larger in 2D than in 3D. This might occur for example if the 3D spiral waves propagating in the disc are subject to a more efficient radiative damping than in 2D discs. In the context of single-star systems, previous work has demonstrated that the amplitude of the spirals depends significantly on the disc cooling timescale (Miranda \& Rafikov 2020a,b; Zhang \& Zhu 2020), and that the AMF may not be conserved under some circumstances. In particular,  spiral waves tend to become weaker as the cooling time increases from a value corresponding to the isothermal limit to $\tcool\sim \Omega^{-1}$. Spiral waves then become stronger again as the cooling time is increased from $\tcool\sim \Omega^{-1}$ to a value corresponding to the adiabatic limit. In adiabatic discs, the AMF is expected to be approximately  constant  (Miranda \& Rafikov 2020a). 

 For our 2D models, the radial profile of the dimensionless cooling time $\beta$ is shown in Fig.~\ref{fig:tcool_vs_r}. We note that it also corresponds to the density-weighted vertically averaged $\tcool$ in our 3D discs.  At the edge of the inner cavity, $\beta\approx 10$  such that the disc behaves almost adiabatically. We therefore expect the AMF associated with waves excited there to be conserved in our 2D simulations. Going back to Fig.~\ref{fig:temp0_k34}, however, we see that for the 3D models, $\beta$ continuously decreases from $\beta\approx 100$ in the equatorial plane to $\beta \approx 10^{-3}$ at the disc surfaces, so that it is not guaranteed that the AMD is conserved in 3D. 
 
 To examine this issue in more detail, we studied the disc response to a perturbing potential of the form:

 \begin{equation}
 \Phi(r,\varphi)= \frac{A}{r^5}\cos(4\varphi-\Omega t),
 \label{eq:pert_pot}
 \end{equation}
 where $A$ is the amplitude of the perturbing potential for which we considered values of $A=0.004$ and 0.04. We note that for this additional set of runs, we used the same thermodynamical treatment as for the simulations of circumbinary discs presented above. The results of these simulations are presented in Fig.~\ref{fig:pert_pot}. In the linear regime with $A=0.004$, we see that the disc response is similar for the 2D and 3D models. In the non-linear regime with $A=0.04$, however, the spiral appears weaker in 3D,  which indicates higher (radiative) dissipation of the spiral wave. In the context of the  simulations of circumbinary discs  presented in Sect.~\ref{sec:results}, this would result in spiral waves with smaller amplitude and consequently in smaller cavity sizes and eccentricities in 3D, in line with what has been observed. A more rapid dissipation of waves in the 3D calculations can be also revealed by  comparing  contours of the surface density in Figs.~\ref{fig:2d_kepler16} and \ref{fig:2d_kepler34}, where we see that spiral waves indeed tend to be stronger in the 2D calculations. 
 
 In summary, it appears that the damping on non-linear spiral waves occurs more rapidly in the 3D models compared to the 2D discs, and the origin of this is the different thermodynamical evolution in 3D versus 2D.

\section{Summary}
In this paper, we have presented the results of three-dimensional hydrodynamical simulations of circumbinary discs around analogues of the Kepler-16(AB) and Kepler-34(AB) binary systems.  We included stellar heating from the central binary, and radiative cooling by employing a cooling term  in which the temperature is relaxed towards the irradiation temperature on a local cooling timescale $\tcool$. The irradiation temperature was calculated by performing a Monte Carlo Radiative Transfer calculation of photon propagation using  RADMC-3D , whereas the cooling timescale $\tcool$ was determining using realistic opacities.

We find that 3D circumbinary discs models show a tendency to reach a quasi-stationary state more rapidly than their 2D counterparts.  Verification of this statement for all models that we have considered is difficult at the present time because of the computational expense of running 3D hydrodynamical simulations, but for a subset of the parameters we have considered the more rapid convergence of 3D simulations versus their 2D counterparts is beyond doubt.
Furthermore, the sizes and eccentricities of the inner cavity are also found to be systematically smaller in 3D models compared to the equivalent 2D models. We interpret this result to be a consequence of the more rapid damping of the non-linear spiral waves that are excited at the cavity edge in 3D versus 2D, or equivalently to be due to the enhanced non-conservation of the angular momentum flux (AMF) associated with these waves as they travel into the disc.
 Although the disc behaves almost adiabatically in our 2D calculations, which would result in a constant AMF for a linear wave that does not undergo non-linear damping, the 3 dimensional structure of the disc contains regions located at intermediate scale heights with $\tcool\sim \Omega^{-1}$, and where spiral wave can be subject to significant radiative damping. Compared to the 2D case, this appears to result in smaller inner cavities that are less eccentric in 3D calculations compared to 2D simulations.
 
A potentially important consequence of these results is that we would expect migrating protoplanets to park closer to the central binary when evolving in a 3D disc model compared to a 2D one, and to maintain a smaller eccentricity. In the case of the circumbinary planet Kepler-34b, this might allow simulations to reproduce the observed orbital parameters of the planet, which has not been possible using 2D simulations to date (e.g. Pierens \& Nelson 2013; Penzlin et al. 2021). Determining whether or not this is the case will be the subject of future work.

\section*{Acknowledgments}
Computer time for this study was provided by the computing facilities MCIA (M\'esocentre de Calcul Intensif Aquitain) of the Universite de Bordeaux and by HPC resources of Cines under the allocation A0110406957 made by GENCI (Grand Equipement National de Calcul Intensif). RPN acknowledges support from Leverhulme Trust through grant number RPG-2018-418 and from STFC through grants ST/P000592/1 and ST/T000341/1.

\begin{appendix}
\section{Effect of viscous heating}

\begin{figure}
\centering
\includegraphics[width=\columnwidth]{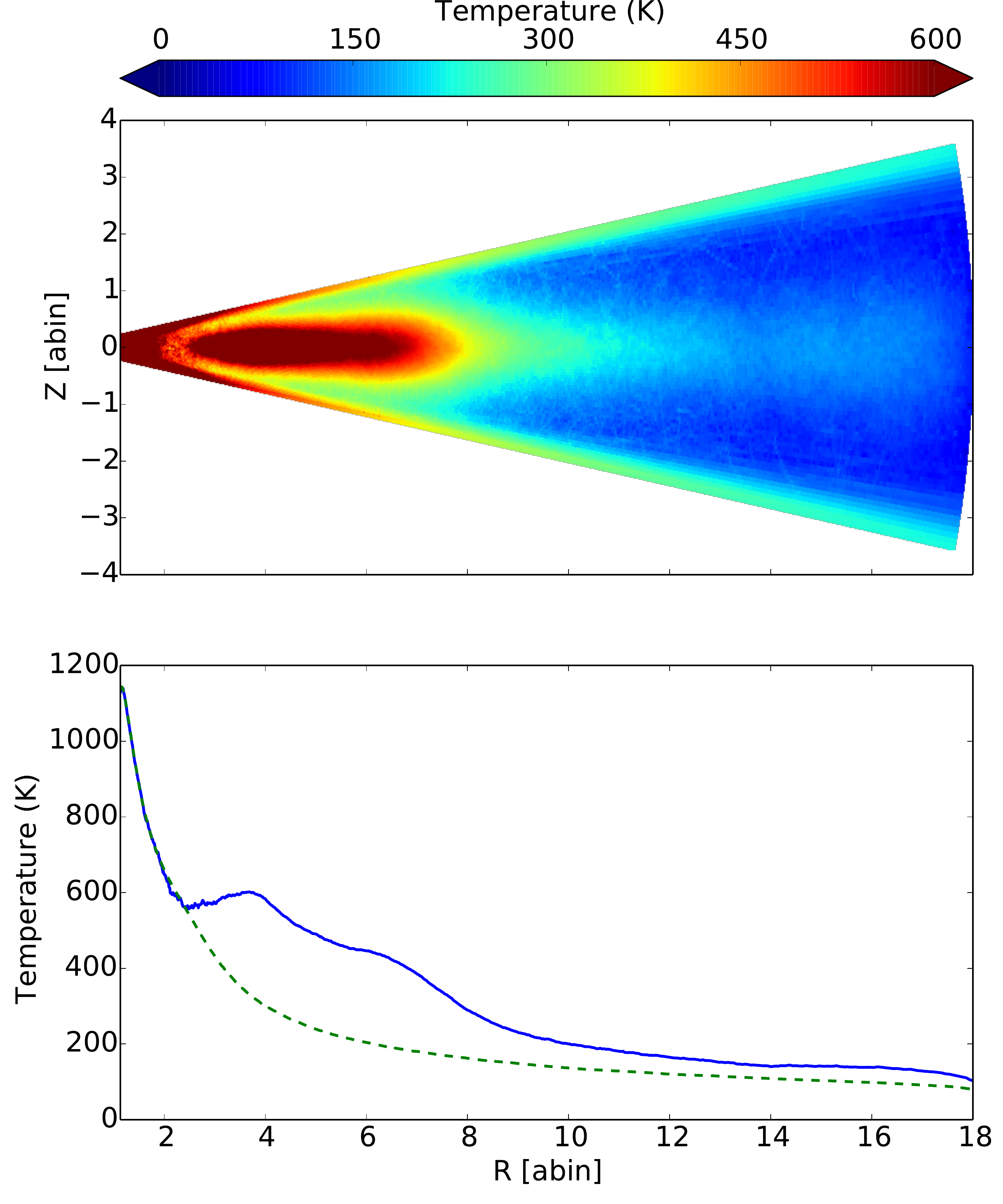}
\caption{Top: for Kepler-34, two-dimensional R-Z temperature distribution  for the initial disc model in the case where the effect of viscous heating is considered. Bottom: corresponding radial temperature profile. Here, we assumed $\alpha=0.01$. }
\label{fig:appendix}
\end{figure}

As mentionned in Sect. \ref{sec:init}, we do not take into account the effect of viscous heating in this study. To estimate the importance of viscous heating relative to stellar heating in a simple way,  we included viscous heating in RADMC-3D and compute the temperature from the initial density field. For $\alpha=0.01$ and binary parameters corresponding to Kepler-34, contours of the outputted temperature are  presented in the top panel of Fig. \ref{fig:appendix}. Comparing this figure with Fig. \ref{fig:temp0_k34}, we see that including viscous heating leads to higher temperatures in the inner regions and in the disc midplane as well, which is not surprising as the strength of viscous heating scales with the value of the gas density. The bottom panel of Fig. \ref{fig:appendix} shows the radial profile of the temperature. Including viscous heating leads to temperatures higher by a factor of $\sim 2$ typically compared to the case of  stellar heating only. This  would result in smaller cavity sizes in simulations where the effect of viscous heating is taken into account.

\end{appendix}

\begin{thebibliography}{}


\bibitem[Artymowicz, \& Lubow(1994)]{1994ApJ...421..651A} Artymowicz, P., \& Lubow, S.~H.\ 1994, ApJ, 421, 651
\bibitem[Bae et al.(2016)]{2016ApJ...833..126B} Bae, J., Nelson, R.~P., \& Hartmann, L.\ 2016, ApJ, 833, 126. doi:10.3847/1538-4357/833/2/126
\bibitem[Bae et al.(2019)]{2019ApJ...884L..41B} Bae, J., Zhu, Z., Baruteau, C., et al.\ 2019, ApJL, 884, L41. doi:10.3847/2041-8213/ab46b0
\bibitem[Barker \& Ogilvie(2014)]{2014MNRAS.445.2637B} Barker, A.~J. \& Ogilvie, G.~I.\ 2014, MNRAS, 445, 2637. doi:10.1093/mnras/stu1939
\bibitem[Ben{\'\i}tez-Llambay, \& Masset(2016)]{2016ApJS..223...11B} Ben{\'\i}tez-Llambay, P., \& Masset, F.~S.\ 2016, ApJS, 223, 11
\bibitem[Bitsch et al.(2013)]{2013A&A...549A.124B} Bitsch, B., Crida, A., Morbidelli, A., et al.\ 2013, A\&A, 549, A124. doi:10.1051/0004-6361/201220159
\bibitem[Coleman et al.(2022)]{2022MNRAS.513.2563C} Coleman, G.~A.~L., Nelson, R.~P., \& Triaud, A.~H.~M.~J.\ 2022, MNRAS, 513, 2563. doi:10.1093/mnras/stac1029
\bibitem[de Val-Borro et al.(2006)]{2006MNRAS.370..529D} de Val-Borro, M., Edgar, R.~G., Artymowicz, P., et al.\ 2006, MNRAS, 370, 529
\bibitem[Doyle et al.(2011)]{2011Sci...333.1602D} Doyle, L.~R., Carter, J.~A., Fabrycky, D.~C., et al.\ 2011, Science, 333, 1602. doi:10.1126/science.1210923
\bibitem[Draine \& Lee(1984)]{1984ApJ...285...89D} Draine, B.~T. \& Lee, H.~M.\ 1984, ApJ, 285, 89. doi:10.1086/162480
\bibitem[Dullemond et al.(2012)]{2012ascl.soft02015D} Dullemond, C.~P., Juhasz, A., Pohl, A., et al.\ 2012, Astrophysics Source Code Library. ascl:1202.015
\bibitem[Flock et al.(2016)]{2016ApJ...827..144F} Flock, M., Fromang, S., Turner, N.~J., et al.\ 2016, ApJ, 827, 144. doi:10.3847/0004-637X/827/2/144
\bibitem[Georgakarakos et al.(2021)]{2021FrASS...8...44G} Georgakarakos, N., Eggl, S., \& Dobbs-Dixon, I.\ 2021, Frontiers in Astronomy and Space Sciences, 8, 44. doi:10.3389/fspas.2021.640830
\bibitem[Holman \& Wiegert(1999)]{1999AJ....117..621H} Holman, M.~J. \& Wiegert, P.~A.\ 1999, AJ, 117, 621. doi:10.1086/300695
\bibitem[Kley \& Haghighipour(2014)]{2014A&A...564A..72K} Kley, W., \& Haghighipour, N.\ 2014, A\& A, 564, A72
\bibitem[Kley, \& Haghighipour(2015)]{2015A&A...581A..20K} Kley, W., \& Haghighipour, N.\ 2015, A\& A, 581, A20
\bibitem[Kley et al.(2019)]{2019A&A...627A..91K} Kley, W., Thun, D., \& Penzlin, A.~B.~T.\ 2019, A\& A, 627, A91. doi:10.1051/0004-6361/201935503
\bibitem[Kostov et al.(2020)]{2020AJ....159..253K} Kostov, V.~B., Orosz, J.~A., Feinstein, A.~D., et al.\ 2020, AJ, 159, 253. doi:10.3847/1538-3881/ab8a48
\bibitem[Kostov et al.(2021)]{2021ApJ...917...93K} Kostov, V.~B., Powell, B.~P., Torres, G., et al.\ 2021, ApJ, 917, 93. doi:10.3847/1538-4357/ac04ad
\bibitem[Li \& Greenberg(1997)]{1997A&A...323..566L} Li, A. \& Greenberg, J.~M.\ 1997, A\& A, 323, 566
\bibitem[Lines et al.(2016)]{2016A&A...590A..62L} Lines, S., Leinhardt, Z.~M., Baruteau, C., et al.\ 2016, A\& A, 590, A62. doi:10.1051/0004-6361/201628349
\bibitem[Lyra et al.(2016)]{2016ApJ...817..102L} Lyra, W., Richert, A.~J.~W., Boley, A., et al.\ 2016, \apj, 817, 102. doi:10.3847/0004-637X/817/2/102
\bibitem[Marzari et al.(2008)]{2008ApJ...681.1599M} Marzari, F., Th{\'e}bault, P., \& Scholl, H.\ 2008, ApJ, 681, 1599. doi:10.1086/588423
\bibitem[Masset et al.(2006)]{2006ApJ...642..478M} Masset, F.~S., Morbidelli, A., Crida, A., et al.\ 2006, ApJ, 642, 478. doi:10.1086/500967
\bibitem[Miranda \& Rafikov(2020)]{2020ApJ...892...65M} Miranda, R. \& Rafikov, R.~R.\ 2020, ApJ, 892, 65. doi:10.3847/1538-4357/ab791a
\bibitem[Miranda \& Rafikov(2020)]{2020ApJ...904..121M} Miranda, R. \& Rafikov, R.~R.\ 2020, ApJ, 904, 121. doi:10.3847/1538-4357/abbee7
\bibitem[Moody et al.(2019)]{2019ApJ...875...66M} Moody, M.~S.~L., Shi, J.-M., \& Stone, J.~M.\ 2019, ApJ, 875, 66. doi:10.3847/1538-4357/ab09ee
\bibitem[M{\"u}ller \& Kley(2012)]{2012A&A...539A..18M} M{\"u}ller, T.~W.~A. \& Kley, W.\ 2012, A\& A, 539, A18. doi:10.1051/0004-6361/201118202
\bibitem[Mu{\~n}oz et al.(2020)]{2020ApJ...889..114M} Mu{\~n}oz, D.~J., Lai, D., Kratter, K., et al.\ 2020, ApJ, 889, 114. doi:10.3847/1538-4357/ab5d33
\bibitem[Mutter et al.(2017)]{2017MNRAS.465.4735M} Mutter, M.~M., Pierens, A., \& Nelson, R.~P.\ 2017, MNRAS, 465, 4735
\bibitem[Penzlin(2021)]{2021A&A...645A..68P} Penzlin, A., Kley, W., Nelson, R.~P. \ 2021, A\&A, 645, 68 
\bibitem[Pierens \& Nelson(2018)]{2018MNRAS.477.2547P} Pierens, A. \& Nelson, R.~P.\ 2018, MNRAS, 477, 2547. doi:10.1093/mnras/sty780
\bibitem[Pierens, \& Nelson(2007)]{2007A&A...472..993P} Pierens, A., \& Nelson, R.~P.\ 2007, A\& A, 472, 993
\bibitem[Pierens \& Nelson(2013)]{2013A&A...556A.134P} Pierens, A., \& Nelson, R.~P.\ 2013, A\& A, 556, A134
\bibitem[Pierens et al.(2020)]{2020MNRAS.496.2849P} Pierens, A., McNally, C.~P., \& Nelson, R.~P.\ 2020, MNRAS, 496, 2849. doi:10.1093/mnras/staa1550
\bibitem[Pierens et al.(2021)]{2021MNRAS.508.4806P} Pierens, A., Nelson, R.~P., \& McNally, C.~P.\ 2021, MNRAS, 508, 4806. doi:10.1093/mnras/stab2853
\bibitem[Shakura \& Sunyaev(1973)]{1973A&A....24..337S} Shakura, N.~I. \& Sunyaev, R.~A.\ 1973, A\&A, 500, 33
\bibitem[Sudarshan et al.(2022)]{2022arXiv220607749S} Sudarshan, P., Penzlin, A.~B.~T., Ziampras, A., et al.\ 2022, arXiv:2206.07749
\bibitem[Tiede et al.(2020)]{2020ApJ...900...43T} Tiede, C., Zrake, J., MacFadyen, A., et al.\ 2020, ApJ, 900, 43. doi:10.3847/1538-4357/aba432
\bibitem[Thun et al.(2017)]{2017A&A...604A.102T} Thun, D., Kley, W., \& Picogna, G.\ 2017, A\& A, 604, A102. doi:10.1051/0004-6361/201730666
\bibitem[Welsh et al.(2012)]{2012Natur.481..475W} Welsh, W.~F., Orosz, J.~A., Carter, J.~A., et al.\ 2012, Nature, 481, 475. doi:10.1038/nature10768
\bibitem[Zhang \& Zhu(2020)]{2020MNRAS.493.2287Z} Zhang, S. \& Zhu, Z.\ 2020, \mnras, 493, 2287. doi:10.1093/mnras/staa404
\bibitem[Zhu et al.(2009)]{2009ApJ...694.1045Z} Zhu, Z., Hartmann, L., \& Gammie, C.\ 2009, ApJ, 694, 1045. doi:10.1088/0004-637X/694/2/1045


\end{thebibliography}
\end{document}